\documentclass[pra,preprint,aps,eqsecnum,showpacs]{revtex4}
\usepackage{graphics}

\begin{document}
\title{Entanglement and nonclassicality for multi-mode radiation field states}
\author{J. Solomon Ivan}
\email{solomon@rri.res.in}
\affiliation{Raman Research Institute, C. V. Raman Avenue, Sadashivanagar, 
Bangalore 560 080, India.}
\author{S. Chaturvedi}

\email{scsp@uohyd.ernet.in}

\affiliation{School of Physics. University of Hyderabad, Hyderabad 500 046, India}

\author{E. Ercolessi}

\email{ercolessi@bo.infn.it}

\affiliation{Physics Dept., University of Bologna, CNISM and INFN, 46 v.Irnerio, I-40126, Bologna, Italy.}

\author{G. Marmo}

\email{marmo@na.infn.it}

\affiliation{Dipartimento di Scienze Fisiche, University of Napoli and INFN, v.Cinzia, I-80126, Napoli, Italy.}

\author{G. Morandi}

\email{morandi@bo.infn.it}

\affiliation{Physics Dept., University of Bologna, CNISM and INFN, 6/2 v.le Berti Pichat, I-40127, Bologna, Italy.}

\author{N. Mukunda}

\email{nmukunda@cts.iisc.ernet.in}

\affiliation{Centre for High Energy Physics, Indian Institute of Science, Bangalore 560 012, India.}

\author{R. Simon}

\email{simon@imsc.res.in}

\affiliation{The Institute of Mathematical Sciences, C.I.T Campus, Chennai 600 113, India.}
\begin{abstract}
Nonclassicality in the sense of quantum optics is a prerequisite for 
entanglement in multi-mode radiation states. In this work we bring out 
the possibilities of passing from the former to the latter, via action 
of   classicality preserving systems like beamsplitters, in a transparent
manner. For single mode states, a complete description of nonclassicality is
available via the classical theory of moments, as a set of necessary and
sufficient conditions on the photon number distribution. We show that when the
mode is coupled to an ancilla in any coherent state, and the system is then
acted upon by a beamsplitter, these conditions turn exactly into signatures of
NPT entanglement of the output state. Since the classical moment problem does
not generalize to two or more modes, we turn in these cases to other familiar
sufficient but not necessary conditions for nonclassicality, namely the Mandel
parameter criterion and its extensions.   We generalize the Mandel matrix from
one-mode states to the two-mode situation, leading to a natural classification
of states with varying levels of nonclassicality. For two--mode states we
present a single test that can, if successful, simultaneously show
nonclassicality as well as NPT entanglement. We also develop a test for NPT
entanglement after beamsplitter action on a nonclassical state, tracing
carefully the way in which it goes beyond the Mandel nonclassicality test. The
result of three--mode beamsplitter action after coupling to an ancilla in the
ground state is treated in the same spirit. The concept of genuine tripartite
entanglement, and scalar measures of nonclassicality at the Mandel level for
two-mode systems, are discussed. Numerous examples illustrating all these
concepts are presented. 

\end{abstract}


\maketitle

\section{Introduction}
States of multi-mode quantized radiation fields are, by very definition,
nonclassical in nature. However for many purposes it is useful to characterize
certain states as displaying nonclassical features in a particularly prominent
or manifest  manner. The quantum optical concept of nonclassicality,  based on
the diagonal coherent state representation \cite{sudarshan63}, and the notion of
entanglement \cite{werner89}, are two such important nonclassical features
displayed by some states but not by others. Whereas the former has played an
important role from the early days of quantum optics \cite{sudarshan63,
stoler70, walls, yuen76, kimdag77, caves81, mandel79, tara92, arvind96,
davidovich96, arvind98}, the latter has received enormous attention more
recently, with the development of the theory of quantum information
\cite{bennett96, simon00,kimble98,  vanloock05, hiro07}. Of course entanglement
is meaningful only in the case of two or more modes of radiation, while
nonclassicality is a useful concept even at the single-mode level.

The main aim of this work is to study the relationships between quantum optical nonclassicality and entanglement of various kinds, and the possibility of generating the latter given the former. In this context the asymmetry between these two features must be carefully appreciated. It is well known that quantum optical nonclassicality is a prerequisite for entanglement \cite{kim02}. Whereas every entangled state is nonclassical, a nonclassical state may be separable (tensor product of nonclassical states, for instance) or entangled. Thus entangled states are a proper subset of nonclassical ones. Then the following questions become meaningful. For every signature of nonclassicality, how much further must one go and what additional conditions have to be met before one is assured of entanglement (of specified type)? Given a nonclassical separable state, can it be transformed into an entangled state through a  physically realizable process using classicality preserving systems?

In speaking in this sense of converting nonclassicality into entanglement we have in mind the use of passive devices such as beamsplitters \cite{campos89, zeilinger94, dutta94} which act on several modes of a given system to produce unitary linear combinations of the annihilation operators of the modes. Such systems preserve total photon number  and, further, they cannot create or destroy nonclassicality. A general study must include developing and clearly specifying signatures of various levels of nonclassicality on the one hand, and of entanglement on the other, and then examining how one can reach the latter starting from the former,  using these classicality preserving passive systems. 

 Two comments by way of clarification are in order before we describe the manner in which the contents of this paper are organized. First, it should be stressed that our concern in this paper is (almost exclusively) with non-Gaussian states. Nonclassicality of Gaussian states is {\underline{necessarily}} of the squeezing type (quadrature squeezing), and a beam splitter can convert such a nonclassicality into entanglement. It is essentially this Gaussian scenario that  largely formed the work-bench during the early years of the theory of entanglement for continuous variable  (canonical) systems, elevating the role of the symplectic group of linear canonical transformations and its unitary (metaplectic) representation. The interplay between nonclassicality and entanglement in the Gaussian case has been long appreciated. For instance, the principal result of Ref.\cite{simon00} can be phrased thus, as emphasised by Kimble recently  \cite{kimble09}: A two-mode (mixed) Gaussian state is separable if and only if its nonclassicality can be removed by local linear canonical transformations --- entanglement is the same as  nonlocal nonclassicality.

 More recently, however, there has emerged considerable interest in non-Gaussian
states \cite{dellanno07, genoni10, dellanno10, auria10, allegra10, missori09,
korolkova08, takahashi09, solomonng08, dellannoPR, ghose07, biswas06}. Ref.\,
\cite{genoni10, dellanno10, dellannoPR} may be consulted for a good introduction
to the literature. Nonclassicality signature of non-Gaussian states is not
restricted to quadrature squeezing --- the state can be antibunched, for
instance. A non-Gaussian state which is neither squeezed nor antibunched could
exhibit nonclassicality signature in the higher moments of the photon count
statistics or photon number distribution \cite{tara92}. Non-Gaussian states thus
exhibit a rich variety of nonclassicality signatures.

 The second comment is in respect of beam splitters; we make extensive use of
this gadget in the present  work. In the domain of quantum optics,
nonclassicality generating operations like squeezing are  considered to be
expensive resources whereas beam splitter, being a passive device, is rightly
considered inexpensive. The latter is passive in the sense that it cannot alter
the total number of photons in the pair of modes it couples, and also in the
stronger sense that it cannot create or destroy nonclassicality.  It thus makes
sense to ask to what extent the expensive  nonclassicality resource can be
converted into the (expensive) entanglement resource using the inexpensive beam
splitter resource.   One should, however, appreciate that the beamsplitter
effects a {\em joint} unitary transformation on the pair of modes it couples,
and hence will be considered an expensive resource in a different context like a
pair of nano mechanical oscillators. Our analysis thus applies to the specific
context of quantum optics.

The contents of this paper are arranged as follows. Section II defines the
concepts of quantum optical classicality --- QO-cl and nonclassicality ---
QO-noncl, for general single-mode fields. In the phase invariant case they are
entirely stated in terms of the photon number distribution \cite{mary97}. This
is possible thanks to the result of the classical Stieltjes moment problem
\cite{shohat}. It is then shown that if such a single-mode state is coupled 
to an ancilla in  any coherent state and passed through any nontrivial $U(2)$
beamsplitter, the resulting two-mode output state shows NPT entanglement
\cite{peres96} precisely when the input single-mode state is quantum optically
nonclassical \cite{asboth05, solo06}. In this case the signatures of the two
coincide exactly. Section III defines quantum optical classicality and
nonclassicality for general two-mode states. We then describe a single test
which if successful is able to establish simultaneously both the nonclassicality
and NPT entanglement of a given two-mode state. Of course proof of entanglement
automatically implies proof of nonclassicality, so the interest here is in the
structure of the expressions used in the test. Since theorems of the moment type
are not available for two or more modes, we describe in Section IV a sufficient
criterion for one-mode nonclassicality originally due to Mandel \cite{mandel79}, and extend
it to two-mode systems as well. All the later considerations of this work are
based on this Mandel level of nonclassicality which is of course weaker than the
general notion of nonclassicality. Several levels or types of nonclassicality in
this sense are described, and three examples to illustrate the ideas are
presented.

In Section V we study states of two-mode systems  for which both notions of
nonclassicality and entanglement make sense. We find conditions for nonclassical
states to become NPT entangled via BS action, and show in detail the way in
which these conditions go beyond those that guarantee Mandel type
nonclassicality. This analysis is applied to two examples to see the formalism
in action. Distillability of the resulting state is demonstrated in one case.
Section VI takes up the theme of Section II and extends it to higher number of
modes. Thus we couple a two-mode state which is nonclassical to a third 
ancilla mode in vacuum, pass this three-mode state through a beamsplitter
acting on the three modes, and develop tests for NPT entanglement in the output
state. As in Section V, here too the precise manner in which testing for
entanglement goes beyond testing for nonclassicality of the Mandel type is
emphasized. Section VII discusses the subtle notion of genuine tripartite
entanglement, while Section VIII brings out some features of two-mode Mandel
level nonclassicality and beamsplitter action. All the ideas in Sections VI,
VII, and VIII are illustrated through several examples with tractable analytical
structures. Section IX contains some concluding remarks. 

\section{Single-mode fields}
Let the photon annihilation and creation operators for the concerned mode be written as $\hat{a}$ and ${\hat{a}}^{\dagger}$, obeying the commutation relation
\begin{equation}
[\hat{a},{\hat{a}}^{\dagger}] \, = 1.
\label{m2.1}
\end{equation}
The familiar Fock states and coherent states are
\begin{eqnarray}
|n \rangle & = & {(n!)}^{-1/2}{({\hat{a}}^{\dagger})}^{n} | 0 \rangle_{}, \,\,\, 
n ~ = ~ 0,1,2, \cdots \;\; ;\nonumber \\
| z_a \rangle & = & e^{-{\frac{1}{2} {|z_a|}^2}} \sum_{n_a=0}^\infty
\frac{z_a^{n_a}}{\sqrt{n_a!}} |n_a \rangle\,,~~
\hat{a}| z_a \rangle  =  z_a | z_a \rangle, \, \,\, z_a \in {\cal C}.
\label{m2.2}
\end{eqnarray}
The Fock states $\{| n \rangle \}$ form an orthonormal basis for the space of
all single-mode states, while the coherent states $\{| z_a \rangle \}$ form a
normalized nonorthogonal overcomplete system.

For a single mode there is no meaning to entanglement, only a separation of
states into the quantum optical classical (QO-cl) and the quantum optical
nonclassical (QO-noncl) types. This is based on the diagonal coherent state
representation of a general (pure or mixed) state $\hat{\rho}^{(a)}$\,:
\begin{equation} 
\hat{\rho}^{(a)}  = {\pi}^{-1}\int_C {d^2 z_a}\, \phi(z_a) | z_a \rangle \langle z_a |. \label{m2.3} 
\end{equation}
The properties of the real diagonal representation weight $\phi(z_a)$ determine
the nature of $\hat{\rho}^{(a)}$ \cite{sudarshan63}\,:
\begin{eqnarray} \phi(z_a) \geq 0 & \Leftrightarrow & \hat{\rho}^{(a)} \,\,
\textrm{QO-cl}; \nonumber \\ \phi(z_a) \not\geq 0 & \Leftrightarrow &
\hat{\rho}^{(a)} \,\, \textrm{QO-noncl}.  \label{m2.4} 
\end{eqnarray}
In the former case, the meaning is that $\phi(z_a)$ is pointwise nonnegative; it
is then mathematically a valid probability density in phase space.

If one is interested only in the expectation values of number conserving
observables, i.e., of operators commuting with $\hat{N}_a = \hat{a}^{\dagger}
\hat{a}$, and hence diagonal in the Fock basis, it suffices to use the phase
averaged form $\hat{\rho}_{D}^{(a)}$ of $\hat{\rho}^{(a)}$\,:

\begin{eqnarray} 
\hat{\rho}_{D}^{(a)} 
& = & \int_{0}^{2 \pi} \frac{d \theta}{2
\pi} e^{i \theta \hat{N}_a} \hat{\rho}^{(a)} e^{-i \theta \hat{N}_a} \nonumber
\\ 
&=& \frac{1}{\pi} \int {d^2 z_a} P(I_a) | z_a \rangle \langle z_a |, \,\,\,
I_a = |z_a|^2, \nonumber \\ 
P(I_a) & = & \int_{0}^{2 \pi} \frac{d \theta}{2 \pi}
\phi(z_a e^{i \theta}) \, , \nonumber \\ 
\hat{\rho}^{(a)}_{D} \hat{N}_a &=&
\hat{N}_a \hat{\rho}^{(a)}_{D}. \label{m2.5} 
\end{eqnarray}
Clearly $\hat{\rho}^{(a)}_{D}$ is a physical state, equivalent to
$\hat{\rho}^{(a)}$ as far as expectation values of number conserving operators
are concerned. Moreover, the quantity $P(I_a)$ is the diagonal representation
weight for $\hat{\rho}^{(a)}_{D}$, so (\ref{m2.4}) leads to the coarse grained
classification \cite{arvind96,arvind98} 
\begin{eqnarray} 
P(I_a) \geq 0 & \Leftrightarrow & {\hat{\rho}}_{D}^{(a)} \,\,
\textrm{QO-cl}; \nonumber \\ 
P(I_a) \not\geq 0 & \Leftrightarrow &
{\hat{\rho}}_{D}^{(a)} \,\, \textrm{QO-noncl}.  \label{m2.6} 
\end{eqnarray}
Further, all information about $P(I_a)$ is contained in the photon number
probabilities 
$p(n)\equiv \langle n|{\hat{\rho}}^{(a)}|n 
\rangle 
=\langle n|{\hat{\rho}}_{D}^{(a)}|n \rangle$ 
for $n =0,1,2, \cdots$ \cite{mary97}. It is convenient to call the set of
probabilities $\{ p(n) \}$ the photon number
distribution (PND). Clearly, $\hat{\rho}^{(a)}_{D}$ can be explicitly
written in terms of the PND and vice versa\,:

\begin{eqnarray}
\hat{\rho}^{(a)}_{D} & = & \sum_{n=0}^{\infty} p(n) | n \rangle
\langle n|, \nonumber \\
p(n)& =& \int_{0}^{\infty} d I_a P(I_a) e^{-I_a} {I_a}^n/ n! \geq 0. 
\label{m2.7}
\end{eqnarray}

Therefore one can ask if the conditions (\ref{m2.6}) for  the coarse grained
QO-cl --- QO-noncl  divide can be  explicitly given in terms of the PND $\{ p(n)
\}$. This is indeed possible  \cite{mary97}, as a result of the classical
analysis of the Stieltjes moment problem. It involves two infinite sequences of
matrix positivity conditions, set up as follows. It is convenient to introduce
the auxiliary quantities
\begin{equation}
q_n = n!p(n), \,\,\,n =0,1,2,\cdots
\label{m2.8}
\end{equation}
Then define two real symmetric infinite dimensional matrices $L$ and
$\tilde{L}$ as follows\,:
\begin{eqnarray}
&& L = (L_{n'n}),\,\,\,\,L_{n'n} = q_{n'+n}, \,\,\,\, n',n = 0,1,2, \cdots ;
\nonumber \\
&&\tilde{L} = (\tilde{L}_{n'n}),\,\,\,\,\tilde{L}_{n'n} = q_{n'+n+1}
, \,\,\,\, n',n = 0,1,2, \cdots  .
\label{m2.9}
\end{eqnarray}
The diagonal elements $L_{nn}$ of $L$ are $q_{2n}$, while $\tilde{L}_{nn}$ of
$\tilde{L}$ are $q_{2n +1}$. Using $L$ and $\tilde{L}$, we can define a sequence
of $(N+1)$-dimensional matrices $L^{(N)}$, ${\tilde{L}}^{(N)}$ as restrictions.
Then the key result is  \cite{mary97}\,:

\begin{eqnarray}
P(I_a) \geq 0 \,\,& \Leftrightarrow & \,\, \{ \hat{\rho}^{(a)}_{D} \,\,
\textrm{and PND}\,\, \{ p(n) \}\,\, \textrm{are QO-cl} \} \nonumber \\
&\Leftrightarrow &\,\,L^{(N)},\,\, \tilde{L}^{(N)} \geq 0, \,\,\,N=0,1,2, \cdots ;
\nonumber \\
L^{(N)} & = & (L_{n'n} \,\,:\,\,n',n=0,1,2,\cdots, N) , \nonumber \\
\tilde{L}^{(N)} & = & (\tilde{L}_{n'n} \,\,:\,\,n',n=0,1,2,\cdots, N).
\label{m2.10} 
\end{eqnarray}
Both $L^{(N)}$ and $\tilde{L}^{(N)}$ are real symmetric $(N+1)$-dimensional
matrices, made up of the intersections of the first $(N+1)$ rows and $(N+1)$
columns of $L$ and $\tilde{L}$ respectively. Conversely we have\,:
\begin{eqnarray}
P(I_a) \not\geq 0 \,\, & \Leftrightarrow & \,\, \{ \hat{\rho}^{(a)}_{D}
\,\,\textrm{and PND}\,\, \{p(n)\}\,\, \textrm{are QO-noncl} \} \nonumber \\
& \Leftrightarrow & \,\, \textrm{some} \,\, L^{(N)}\not\geq 0\,\, 
\textrm{and or some} \,\, \tilde{L}^{(N)} \not\geq 0 .
\label{m2.11} 
\end{eqnarray}

As noted earlier, for one-mode states there is no concept of entanglement.
However a QO-noncl state characterized by (\ref{m2.11}) can be converted to a
state possessing NPT entanglement by adjoining it to a second  ancilla mode, the
$b$-mode, initially say in its vacuum state, and passing this two-mode state
through a generic beamsplitter (BS) corresponding to an element of the group
$U(2)$. This is seen as follows\,:

The $b$-mode brings in an added operator pair $\hat{b}$, $\hat{b}^{\dagger}$
obeying the same commutation relation (\ref{m2.1}); and each of $\hat{a}$,
$\hat{a}^{\dagger}$ commutes with $\hat{b}$ and $\hat{b}^{\dagger}$. Let now  
\begin{equation}
u =
\left( \begin{array}{cc}
u_{11} & u_{12} \\
u_{21} & u_{22}
\end{array} \right)
\label{m2.12}
\end{equation}
be a general element of the two-dimensional unitary group $U(2)$. The action of
the corresponding BS on the operators of the two modes is given by conjugation
with a unitary operator $\hat{U}$ acting on the two-mode Hilbert space, the
dependence of $\hat{U}$ on $u$ being left implicit \cite{campos89, arvind96}\,:
\begin{eqnarray}
\hat{U} \left( \begin{array}{c}
\hat{a} \\
\hat{b} \end{array} \right) {\hat{U}}^{-1} 
= u^{\dagger}
\left( \begin{array}{c}
\hat{a} \\
\hat{b} \end{array} \right), &&
\hat{U} \left( \begin{array}{c}
\hat{a}^{\dagger} \\
\hat{b}^{\dagger} \end{array} \right) {\hat{U}}^{-1} 
= u^{T}
\left( \begin{array}{c}
\hat{a}^{\dagger} \\
\hat{b}^{\dagger} \end{array} \right) ; \nonumber \\ 
{\hat{U}}^{-1} \left( \begin{array}{c}
\hat{a} \\
\hat{b} \end{array} \right) \hat{U} 
= u
\left( \begin{array}{c}
\hat{a} \\
\hat{b} \end{array} \right), &&
{\hat{U}}^{-1} \left( \begin{array}{c}
\hat{a}^{\dagger} \\
\hat{b}^{\dagger} \end{array} \right) \hat{U} 
= u^{*}
\left( \begin{array}{c}
\hat{a}^{\dagger} \\
\hat{b}^{\dagger} \end{array} \right).
\label{m2.13}
\end{eqnarray}
This unitary operator is `passive' in the sense that 
\begin{equation}
\hat{U} (\hat{N}_a + \hat{N}_b ) = (\hat{N}_a + \hat{N}_b ) \hat{U}.
\label{m2.14}
\end{equation}
It is also passive in another important sense in the present context\,: for any
two-mode state the property of being QO-cl or QO-noncl is preserved upon passage
through any BS, so BS action can neither create nor destroy nonclassicality. (
The output state of a beam splitter is a coherent state if and only if the input
is. This will be clarified and elaborated in the next Section).

We start with the $a$-mode state $\hat{\rho}^{(a)}_{D}$ which commutes with
$\hat{N}_a$, and take as the two-mode input to the BS the
separable (product) state
\begin{equation}
\hat{\rho}_{\rm in}^{(ab)} = \hat{\rho}^{(a)}_{D} \otimes |0 \rangle_b {}_b 
\langle 0|.
\label{m2.15}
\end{equation}
This commutes with the total number operator $\hat{N}_a + \hat{N}_b$. Passage
through the BS preserves this property and results in the output state
\begin{eqnarray}
\hat{\rho}^{(ab)}_{\rm out} & = & \hat{U} \hat{\rho}^{(ab)}_{\rm in} {\hat{U}}^{-1} \nonumber \\
& = & \hat{U} \sum_{n=0}^{\infty} \frac{p(n)}{n!} (\hat{a}^{\dagger})^n
|0,0 \rangle \langle 0, 0| (\hat{a})^n {\hat{U}}^{-1} \nonumber \\
& = &\sum_{n=0}^{\infty} \frac{p(n)}{n!} (u_{11} \hat{a}^{\dagger} + 
u_{21} \hat{b}^{\dagger} )^n
|0,0 \rangle \langle 0, 0| (u_{11}^{*}\hat{a}+ u_{21}^{*}\hat{b} )^n.
\label{m2.16}
\end{eqnarray}
As a general notation let us now use $n, n',n'',\cdots$ to denote the
eigenvalues of $\hat{N}_a$, and $m, m', m'', \cdots$ to denote eigenvalues of
$\hat{N}_b$. Then the general Fock state matrix elements of
$\hat{\rho}^{(ab)}_{\rm out}$ are\,: 
\begin{equation}
\langle n' , m' |\hat{\rho}^{(ab)}_{\rm out} | n ,m \rangle =
\delta_{n'+ m',n+m}\, q_{n+m}\, \frac{u_{11}^{n'} u_{21}^{m'}}{\sqrt{n'! m'!}}
\frac{u_{11}^{*n} u_{21}^{*m}}{\sqrt{n! m!}}.
\label{m2.17}
\end{equation}
To see whether $\hat{\rho}^{(ab)}_{\rm out}$ possesses NPT entanglement, we
carry out the partial transpose (PT) operation by implementing it 
in the $b$-mode space  in the Fock basis. This amounts to 
interchanging 
$m$ and $m'$ on the right hand side of Eq. (\ref{m2.17}), and gives\,:
\begin{equation}
\langle n' , m' |\hat{\rho}^{(ab) PT}_{\rm out} | n ,m \rangle =
\delta_{n'+ m,n+m'}\, q_{n+m'}\, \frac{u_{11}^{n'} u_{21}^{m}}{\sqrt{n'! m!}}
\frac{u_{11}^{*n} u_{21}^{*m'}}{\sqrt{n! m'!}}.
\label{m2.18}
\end{equation}
 The question now is whether this is a physical state i.e., if $\hat{\rho}^{(ab) PT}_{\rm out} \geq 0$

We now isolate two principal submatrices out of the matrix ${\hat{\rho}}^{(ab)PT}_{\rm out}$ in the Fock basis, 
which are closely related to the matrices $L$, $\tilde{L}$ of Eq. (\ref{m2.9}). The first one is the submatrix $H=(H_{n'n})$ obtained by taking $m'=n'$, $m=n$ in (\ref{m2.18})\,:
\begin{eqnarray}
H_{n'n} &=&\langle n',n'|\hat{\rho}_{\rm out}^{(ab)PT}|n,n\rangle 
\nonumber \\
& = & q_{n+n'} \frac{{(u_{11}^{} u_{21}^{*})^{n'}}}{n'!}
\frac{(u_{11}^{*} u_{21}^{})^n}{{n!}} \, , \,\,\, n', n = 0, 1, 2, \cdots;
\nonumber \\
H &=& A^{\dagger} L A, \,\,A = {\rm diag}(\frac{(u_{11}^{*} u_{21})^{n}}{{n!}}
,\,\,n=0,1,2, \cdots).
\label{m2.19}
\end{eqnarray}
The second is the submatrix $\tilde{H} = (\tilde{H}_{n'n})$ obtained by
taking $m' = n'+1$, $m=n+1$ in (2.18)\,:
\begin{eqnarray}
\tilde{H}_{n'n}&=&\langle n',n'+1|\hat{\rho}_{\rm out}^{(ab)PT}|n,n+1\rangle 
\nonumber \\
&=& q_{n+n'+1} \frac{{(u_{11}^{} u_{21}^{*})^{n'}}}{n'!}
\frac{u_{21}^{*}}{\sqrt{n'+1}}
\frac{(u_{11}^{*} u_{21}^{})^n}{{n!}}\frac{u_{21}^{}}{\sqrt{n+1}}
 \, , \,\,\, n', n = 0, 1, 2, \cdots;
\nonumber \\
\tilde{H} &=& \tilde{A}^{\dagger} \tilde{L} \tilde{A}, \,\,
\tilde{A} = {\rm diag}(\frac{(u_{11}^{*} u_{21}^{})^{n}}{{n!}}
\frac{u_{21}^{}}{\sqrt{n+1}}
,\,\,n=0,1,2, \cdots).
\label{m2.20}
\end{eqnarray}
Invertibility of $A$ and $\tilde{A}$ implies that $H,\,\tilde{H} \ge 0$ if and
only if $L,\,\tilde{L}\ge 0$.  We then see: every signature of
$\hat{\rho}_{D}^{(a)}$ and the PND $\{ p(n) \}$ being QO-noncl, such as some
$L^{(N)} \not\geq 0$ or some $\tilde{L}^{(N)} \not\geq 0$, directly implies a
corresponding signature of $\hat{\rho}_{\rm out}^{(ab)}$ being NPT entangled,
since either $H \not\geq 0$ or $\tilde{H} \not\geq 0$, implying that
$\hat{\rho}_{\rm out}^{(ab)PT} \not \geq 0$.   It is in this precise sense that
any nontrivial BS is able to convert input QO-nonclassicality of a single-mode
to NPT entanglement of the output state $\hat{\rho}_{\rm out}^{(ab)}$, in a
sense preserving the signature of nonclassicality. 

A simple calculation shows that in the input state (\ref{m2.15}) of the $ab$-system, 
the $b$-mode need not be in the vacuum but could be in a general coherent state 
$| z_b\rangle_b$ for some complex nonzero $z_b$. Since 
\begin{eqnarray}
| z_b\rangle_b & = & D^{(b)}(z_b) | 0 \rangle_b , \nonumber \\
D^{(b)}(z_b) & = & \exp (z_b \hat{b}^{\dagger} - z_{b}^{*} \hat{b}),
\label{m2.21}
\end{eqnarray}  
we have the following replacements for the previous Eqs. (\ref{m2.15}, \ref{m2.16})\,:
\begin{eqnarray}
\hat{\rho}_{\rm in}^{(ab)} &=& \hat{\rho}^{(a)}_{D} \otimes |z_b \rangle_b {}_b 
\langle z_b| \nonumber \\
& = &D^{(b)}(z_b) \{ \hat{\rho}^{(a)}_{D} \otimes |0 \rangle_b {}_b\langle 0| \} {D^{(b)}
(z_b)}^{-1} ; \nonumber \\
\hat{\rho}_{\rm out}^{(ab)} &=& \hat{U} \hat{\rho}_{\rm in}^{(ab)} {\hat{U}}^{-1} \nonumber \\
& = &D^{(a)}(u_{12} z_b)D^{(b)}(u_{22} z_b) \{ \hat{U} \hat{\rho}^{(a)}_{D} \otimes 
|0 \rangle_b {}_b\langle 0|{\hat{U}}^{-1} \} {D^{(a)}(u_{12} z_b)}^{-1} {D^{(b)}
(u_{22} z_b)}^{-1}.
\label{m2.22}
\end{eqnarray}
(In passing we note that this input state no longer commutes with
$\hat{N}_{a}+\hat{N}_{b}$). Here we have used Eq. (\ref{m2.13}). Thus the matrix
elements of this two-mode BS output state in the displaced orthonormal product
basis $D^{(a)}(u_{12} z_b)D^{(b)}(u_{22} z_b) | n, m \rangle$ are identically
the same as the matrix elements of $\hat{\rho}_{out}^{(ab)}$ of Eq.
(\ref{m2.16}) in the Fock basis $| n, m \rangle$. Implementing the PT operation
in this new product basis we recover the earlier results. Thus every signature
of QO-noncl of $\hat{\rho}^{(a)}_{D}$ is transformed by a general BS action into
a corresponding signature of NPT entanglement of the output two-mode state
(\ref{m2.22}).

In both cases (\ref{m2.16}) and (\ref{m2.22}) we see that nontrivial BS action,
while being passive and so maintaining any QO nonclassicality present in the
input state, is able to convert an initially unentangled two-mode state into an
NPT entangled state. This it can do provided there is some QO nonclassicality to
begin with. Crudely speaking, what the beam splitter achieves in the present
case is to take the input  nonclassicality which resides `locally' in the
$a$-mode and convert it into nonclassicality residing `nonlocally' as
entanglement between the modes. 

\section{Two-mode fields - General properties
and an entanglement test}
We now consider specific new features encountered in the study of states of a
two-mode system. A general two-mode state $\hat{\rho}^{(ab)}$ possesses the
diagonal coherent state representation
\begin{equation}
\hat{\rho}^{(ab)}  =   \int \int \frac{d^2 z_a}{\pi}\, \frac{d^2 z_b}{\pi}\, \phi(z_a, z_b) 
| z_a, z_b \rangle \langle z_a, z_b |
\label{m3.1}
\end{equation}
in terms of the two-mode (product) coherent states $| z_a ,z_b \rangle$.
Analogous to Eq. (\ref{m2.4}), the properties of $\phi(z_a, z_b)$ determine the
nature of $\hat{\rho}^{(ab)}$\,: 
\begin{eqnarray}
\phi(z_a, z_b) \geq 0 & \Leftrightarrow & \hat{\rho}^{(ab)} \,\, \textrm{QO-cl}; \nonumber \\
\phi(z_a,z_b) \not\geq 0 & \Leftrightarrow & \hat{\rho}^{(ab)} \,\, \textrm{QO-noncl}.  
\label{m3.2}
\end{eqnarray}
In the former case $\phi(z_a, z_b)$ has the properties of a
probability distribution in the four dimensional phase space.

Now, apart from examining whether a given state $\hat{\rho}^{(ab)}$ is QO-cl or
QO-noncl, we can also ask in the latter case whether it is entangled, and if so
whether it is NPT type, distillable \cite{horodecki97}, etc. These added
questions become meaningful. In fact we will develop later in this Section an
interesting test or criterion which can witness simultaneously for QO-noncl
assicality as well as NPT entanglement. 

With respect to BS action (\ref{m2.12},
\ref{m2.13}) representing general elements $u$ $\in$ $U(2)$, we note the
following. Such action is nonlocal since the modes $a$ and $b$ get linearly
mixed, in addition to being passive in the sense of conserving $\hat{N}_{a}+
\hat{N}_{b}$. Since annihilation operators go into linear combinations of
annihilation operators under this action, coherent states go into coherent
states with the matrix $u$ acting on the pair $z_a$, $z_b$ as a column vector.
Therefore convex sums of coherent states go into similar convex sums
\cite{aharonov66},   $\phi(z_a, z_b)$ experiences a point transformation  
$\phi(z_a, z_b) \to \phi(u_{11}z_a + u_{12}z_b,\, u_{21}z_a +u_{22}z_b)$, and
thus such BS action preserves the QO-cl or QO-noncl nature of the state
$\hat{\rho}^{(ab)}$. This is another sense of BS action being passive.  

On the other hand, while a QO-cl state has no entanglement, a QO-noncl state may
be separable, i.e., unentangled, or may possess entanglement. Entangled states
are a proper subset of QO-noncl states, and NPT entangled states are a further
subset. Now BS action can cause a transition, within the QO-noncl subset, from a
separable to an entangled state, in which case we can further enquire into the
nature of the entanglement so obtained, whether it is NPT type etc. This is in
fact the case in the transition from $\hat{\rho}^{(ab)}_{\rm in}$ of Eq.
(\ref{m2.15}) to $\hat{\rho}_{\rm out}^{(ab)}$ of Eq. (\ref{m2.16}), or the
transition in Eq. (\ref{m2.22}).

Continuing with the definitions of QO-cl and QO-noncl in Eq. (\ref{m3.2}), if we
are interested only in the total number conserving matrix elements of various
operators, i.e., of operators commuting with $\hat{N}_a + \hat{N}_b$, it
suffices to work with the total phase averaged state 
\begin{eqnarray}
\hat{\rho}_{D}^{(ab)} & =&  \int_{0}^{2 \pi} \frac{d \theta}{2 \pi}
e^{i \theta (\hat{N}_a +\hat{N}_b)} \hat{\rho}^{(ab)} e^{-i \theta (\hat{N}_a +\hat{N}_b)}
\nonumber \\
& = &  \int \frac{d^2 z_a}{\pi}\frac{d^2 z_b}{\pi} P(I_a, I_b, \theta) 
| z_a, z_b \rangle \langle z_a, z_b |, \nonumber \\
I_a & =& |z_a|^2,\,\,\,I_b = |z_b|^2,\,\,\,\theta = {\rm arg} z_{a}^{*}z_b ,
 \nonumber \\
P(I_a , I_b, \theta) & = &  \int_{0}^{2 \pi} \frac{d \theta'}{2 \pi} 
\phi(\sqrt{I_a}e^{-i \theta'} ,\sqrt{I_b}e^{i (\theta- \theta')} ).
\label{m3.3}
\end{eqnarray}
This state is number conserving\,:
\begin{eqnarray}
\hat{\rho}_{D}^{(ab)} (\hat{N}_a + \hat{N}_b) &=& (\hat{N}_a + \hat{N}_b)
\hat{\rho}_{D}^{(ab)} , \nonumber \\
\langle n'  m' |\hat{\rho}_{D}^{(ab)} | n m \rangle & = & 
\delta_{n'+m',n+m} \langle n'  m' |\hat{\rho}^{(ab)} | n m \rangle. 
\label{m3.4}
\end{eqnarray}
Since  the coarse grained  $P(I_a , I_b, \theta)$ is the (real) diagonal
representation weight of $\hat{\rho}_{D}^{ab}$, we have the following QO
classification at this level \cite{arvind96,arvind98}\,: 
\begin{eqnarray}
&&P(I_a , I_b, \theta) \geq 0 \,\, \Leftrightarrow \,\,\hat{\rho}_{D}^{(ab)}
\,\,\textrm{is QO-cl}, \nonumber \\
&&P(I_a , I_b, \theta) \not\geq 0 \,\, \Leftrightarrow \,\,\hat{\rho}_{D}^{(ab)}
\,\,\textrm{is QO-noncl}.
\label{m3.5} 
\end{eqnarray}
We use this in the next Section.

We now revert to a general state $\hat{\rho}^{(ab)}$ and describe a test which,
if it succeeds, {\em simultaneously establishes} both QO nonclassicality of
$\hat{\rho}^{(ab)}$ and its NPT entanglement. It is clear of course, that any
test which establishes the latter immediately implies the former; the interest
here is in the structure of the test itself.

We set up an infinite matrix $\hat{N}$ with operator entries $\hat{N}_{jk,lm}$
where $j,k,l,m$ run independently over $0,1,2,\cdots$. The pair $jk$ denotes a
`row index' and takes in sequence the values $00;10,01;20,11,02;30,21,12,03;
\cdots$ . Similarly the `column index' pair $lm$ also takes these same values in
the same sequence. We define the entries of $\hat{N}$ thus\,:
\begin{eqnarray}
&& \hat{N}_{jk,lm} ={\hat{N}_{lm,jk}}^{\,\,\,\dagger} = \hat{a}^{\dagger j}
\hat{b}^{\dagger k} \hat{a}^{l} \hat{b}^m.
\label{m3.6}
\end{eqnarray}
Clearly $\hat{N} = \left( \hat{N}_{jk,lm} \right)$ is an infinite `hermitian'
matrix of operator entries. Note that these entries are in normal-ordered form.
Starting with the diagonal representation (\ref{m3.1}), for any set of complex
coefficients $\{c_{jk} \}$ and the associated positive semidefinite operator
$\sum_{jk,lm} c_{jk}^{*}\hat{N}_{jk,lm} c_{lm}$, we always have\,:
\begin{eqnarray}
{\rm Tr}(\hat{\rho}^{(ab)} \sum_{jk,lm} c_{jk}^{*}\hat{N}_{jk,lm} c_{lm})
& =& {\rm Tr}(\hat{\rho}^{(ab)} (\sum_{jk} c_{jk}\hat{a}^{j} 
\hat{b}^k)^{\dagger}
 (\sum_{lm} c_{lm}\hat{a}^{l} \hat{b}^m)) \nonumber \\
& =& \int \int \frac{d^2 z_a}{\pi} \, \frac{d^2 z_b}{\pi}\,\phi(z_a, 
z_b) 
|\sum_{lm} c_{lm} z_{a}^{l} z_{b}^{m} |^2 \geq 0.
\label{m3.7}
\end{eqnarray}
This is  independent of $\phi(z_a, z_b)$ being classical or otherwise, because
we have here the expectation value of a positive semidefinite hermitian
operator. On the other hand, if we pass to the partial transpose
$\hat{\rho}^{(ab) PT}$ of $\hat{\rho}^{(ab)}$, by performing transposition only
in the space of states of the $b$-mode in the Fock basis, this will amount to
everywhere replacing $\hat{b}^{\dagger j}\hat{b}^m$ by $\hat{b}^{\dagger
m}\hat{b}^j$, since in the Fock basis $\hat{b}^{\dagger}$ and $\hat{b}$ are real
\cite{vogel05}. Thus for the same positive semidefinite operator as in
(\ref{m3.7}) we have\,:
\begin{eqnarray}
 {\rm Tr}(\hat{\rho}^{(ab) PT} \sum_{jk,lm} c_{jk}^{*}\hat{N}_{jk,lm} c_{lm})
& =& {\rm Tr}(\hat{\rho}^{(ab)} \sum_{jk,lm} c_{jk}^{*}\hat{a}^{\dagger j} 
\hat{b}^{\dagger m}
 \hat{a}^{l} \hat{b}^k c_{lm}) \nonumber \\
& =& \int \int \frac{d^2 z_a}{\pi} \, \frac{d^2 z_b}{\pi}\, \phi(z_a, 
z_b)\, 
|\sum_{lm} c_{lm} z_{a}^{l} z_{b}^{*m} |^2.
\label{m3.8}
\end{eqnarray}
Notice the difference in the integrands of the last integrals in (\ref{m3.7})
and (\ref{m3.8}); the latter integral is sure to be positive if
$\hat{\rho}^{(ab) PT}  \geq 0$,  otherwise it could be negative.

Thus we arrive at a single step test for QO-nonclassicality and NPT entanglement
of $\hat{\rho}^{(ab)}$. The above expression (\ref{m3.8}) being negative implies
{\em two things simultaneously}\,: 
\begin{eqnarray}
&& {\rm (i)} \,\,\phi(z_a, z_b) \not\geq 0, \,\,{\rm hence} \,\,\hat{\rho}^{(ab)} \,\,
{\rm is \,\, QO}-{\rm noncl} ; \nonumber \\
&& {\rm (ii)}\,\, \hat{\rho}^{(ab)PT} \not\geq 0, \,\,\, \textrm{and hence}\,\,
\hat{\rho}^{(ab)} \,\,\textrm{ is NPT entangled}.
\label{m3.9}
\end{eqnarray}
As we said already, $\hat{\rho}^{(ab)}$ being NPT entangled already
implies also its being QO-noncl. The point here is that the expression
(\ref{m3.8}) being negative manifestly displays both properties of
$\hat{\rho}^{(ab)}$ immediately.

This interesting result is an indication of the possibility, in
suitable circumstances, of `bridging the gap' between the
characterization of QO nonclassicality and the further
characterization of (NPT) entanglement for two-mode fields in an
efficient manner. For the particular the test based on (\ref{m3.8}), there
is no `gap' at all, but as we will show this will not always be the case.     

\section{Mandel matrices and state classification for one and two-mode
fields}

The discussion of single-mode states in Section II made use of the probabilities
$p(n)$, $n=0,1,2,\cdots,$ individually. In terms of the moments of the PND, this
means in principle that all its moments are involved. It is naturally possible
to make much more limited statements about QO nonclassicality if one uses only
the first and second moments, say, of the PND. This is the content of the Mandel
criterion for QO nonclassicality stated in terms of the Mandel matrix (or Mandel
$Q$ parameter) associated with a given state. It is useful to begin by outlining
this for single-mode states to set up notations, and then generalize to two
modes.

Another motivation is that for two or more modes there are no moment theorems at
all comparable in scope to the one stemming from the classical Stieltjes moment
problem, and therefore no  simple generalizations of Eqs. (\ref{m2.10},
\ref{m2.11}) into necessary and sufficient conditions for QO classicality   can
be expected to be available.

\subsection{Mandel matrices for single-mode states}
Given a one-mode state $\hat{\rho}^{(a)}$, we construct by Eq. (\ref{m2.5}) the
state $\hat{\rho}^{(a)}_{D}$ conserving $\hat{N}_a$. The fact that
$\hat{\rho}^{(a)}_{D}$ is a physical state leads to a certain $2 \times 2$ real
symmetric matrix, involving the first and second moments of the PND $\{ p(n)
\}$, being positive semidefinite\,:
\begin{eqnarray}
\hat{\rho}^{(a)}_{D} & \geq & 0 \Rightarrow 
\left( \begin{array}{cc}
1 & \langle\hat{N}_a \rangle \\
\langle\hat{N}_a \rangle & \langle\hat{N}_{a}^{2} \rangle 
\end{array} \right)
=\left( \begin{array}{cc}
 1 & \langle \hat{a}^{\dagger} \hat{a}\rangle \\
\langle \hat{a}^{\dagger} \hat{a}\rangle &
\langle \hat{a}^{\dagger} \hat{a}\hat{a}^{\dagger} \hat{a}\rangle
\end{array} \right) \geq 0 , \nonumber \\
\langle\hat{N}_a \rangle & = & \sum_{n=0}^{\infty} n p(n), \,\,\,\,
\langle\hat{N}_{a}^{2} \rangle =  \sum_{n=0}^{\infty} n^2 p(n).
\label{m4.1}
\end{eqnarray} 
All expectation values here are in the state $\hat{\rho}^{(a)}_{D}$.
The Mandel matrix associated with $\hat{\rho}^{(a)}_{D}$ is obtained by replacing 
the expectation value $\langle\hat{N}_{a}^{2} \rangle$ by the 
expectation value of the normal ordered form $\hat{a}^{\dagger 2}
\hat{a}^{2}$ of $\hat{N}_{a}^{2} $\,:
\begin{equation}
M^{(1)}(\hat{\rho}^{(a)}_{D}) =
\left( \begin{array}{cc}
 1 & \langle \hat{a}^{\dagger} \hat{a}\rangle \\
\langle \hat{a}^{\dagger} \hat{a}\rangle &
\langle \hat{a}^{\dagger 2} \hat{a}^{2} \rangle
\end{array} \right).
\label{m4.2}
\end{equation}
Here the superscript ${}^{(1)}$ indicates that we are considering states of a
single-mode. While positivity of $\hat{\rho}^{(a)}_{D}$ implies the positivity
of the $2 \times 2$ matrix in (\ref{m4.1}), it does not imply the positivity of
$M^{(1)}(\hat{\rho}^{(a)}_{D})$. From Eq. (\ref{m2.7}) we easily obtain\,:
\begin{eqnarray}
 \langle\hat{a}^{\dagger} \hat{a}\rangle &=& \sum_{n=0}^{\infty} n p(n) = 
\int_{0}^{\infty} d I_a P(I_a) I_a = \langle I_a \rangle, \nonumber \\
\langle \hat{a}^{\dagger} \hat{a}\hat{a}^{\dagger} \hat{a} \rangle &=&
\sum_{n=0}^{\infty} n^2 p(n) = \int_{0}^{\infty} d I_a P(I_a) I_a(I_a 
+1)= \langle I_a (I_a +1) \rangle,
\nonumber \\
\langle \hat{a}^{\dagger 2} \hat{a}^{2} \rangle &=&  
\sum_{n=0}^{\infty} n(n-1) p(n) = \int_{0}^{\infty} d I_a P(I_a) 
I_{a}^{2}= \langle I_{a}^{2} \rangle,
\nonumber \\
\langle \hat{a}^{\dagger 2} \hat{a}^{2} \rangle -
\langle\hat{a}^{\dagger} \hat{a}\rangle {}^2 &=&
\int_{0}^{\infty} d I_a P(I_a) (I_{a} -\langle I_a \rangle)^{2}.  
\label{m4.3}
\end{eqnarray}
Clearly, the last expression cannot be negative if $P(I_a)$ is pointwise
nonnegative. Thus we are led to the Mandel classification of states of a
single-mode field \cite{mandel79, arvind98}\,:
\begin{eqnarray*}
 P(I_a ) \geq 0, \,\,\hat{\rho}^{(a)}_{D}\,\,\textrm{is QO-cl}
&\Rightarrow& \,\,
\{ M^{(1)}(\hat{\rho}^{(a)}_{D}) \geq 0 \Leftrightarrow {\rm det}
M^{(1)}(\hat{\rho}^{(a)}_{D}) \geq 0  \nonumber \\
& \Leftrightarrow& (\Delta N_a)^2 - \langle N_a \rangle \geq 0\}: \nonumber \\
&&\hat{\rho}^{(a)}_{D} \,\,\textrm{displays super-Poissonian statistics
(super-PS)};\nonumber 
\end{eqnarray*}

\begin{eqnarray}
\{ M^{1}(\hat{\rho}^{(a)}_{D}) \not\geq 0 &\Leftrightarrow& {\rm det}
M^{(1)}(\hat{\rho}^{(a)}_{D}) < 0
\Leftrightarrow (\Delta N_a)^2 - \langle N_a \rangle < 0 \}
\Rightarrow \nonumber \\
P(I_a) &\not\geq& 0, 
\,\,\hat{\rho}^{(a)}_{D}\,\, \textrm{is QO-noncl}\,:\nonumber \\
&&\,\hat{\rho}^{(a)}_{D} \,\,\textrm{displays sub-Poissonian statistics
(sub-PS)}.
\label{m4.4}
\end{eqnarray}
It is because the Mandel matrix is two dimensional with obviously
positive trace that the positivity or nonpositivity of 
$M^{(1)}(\hat{\rho}^{(a)}_{D})$ reduces to that of its determinant, hence to
that of $(\Delta N_a)^2 - \langle N_a \rangle$.

\subsection{Single-mode squeezed vacuum example}
As a useful and instructive example of this concept, we consider the case of a
single-mode  squeezed  vacuum \cite{stoler70, yuen76, caves81}. Such a state is
obtained by applying a unitary (scaling) operator involving the exponential of a
complex combination of $\hat{a}^{{\dagger 2}}$ and $\hat{a}^{2}$ to the Fock
vacuum $|0 \rangle_a$, and is parametrised by a complex variable $\xi = \xi_1 +
i\xi_2$ or an equivalent complex variable $\omega$\,:
\begin{eqnarray}
|\psi^{(a)} (\omega) \rangle &=& \exp \{{\frac{1}{4} (\xi \hat{a}^{\dagger 2} - \xi^{*}
\hat{a}^2)}\} | 0 \rangle_a \nonumber \\
&=& (1-|\omega|^2)^{\frac{1}{4}} 
\sum_{n=0}^{\infty} \sqrt{\frac{\Gamma (n+1/2)}{n!\sqrt{\pi}}
}\, {\omega}^{n}\, | 2n \rangle_a , \nonumber \\
\omega &=& \frac{\xi}{|\xi|} {\rm tanh}{(|\xi|/2)}.
\label{m4.5}
\end{eqnarray}
Since only even photon number states are present, the probabilities $p(1)$,
$p(3)$, $p(5)$, $\cdots$ in the PND vanish.  That is, $\tilde{L} \ne 0$   but
all its diagonals vanish, implying  $\tilde{L}\not\geq 0$,  which is immediate
evidence that these states are QO-noncl. Some important expectation values
are\,:
\begin{eqnarray}
&&\langle \psi^{(a)} (\omega) | \{ \hat{a}^{\dagger}, \hat{a},\,\, 
\hat{N}_{a},\,\,
\hat{N}_{a}^{2},\,\, \hat{a}^{\dagger 2} \hat{a}^2,\,\, 
\hat{a}^2 \}| \psi^{(a)} (\omega) \rangle = \nonumber \\
&&~~~~~~~~~~~~~~~\{ 0,\,\, 0,\,\, S^2,\,\, S^2(S^2+2C^2),\,\, S^2(2S^2+C^2),\,\, 
\frac{\xi}{|\xi|} \},\nonumber \\
&&~~~~S = {\rm sinh}(|\xi|/2) , \,\, C = {\rm cosh}(|\xi|/2).
\label{m4.6}
\end{eqnarray}
The $2\times 2$ Mandel matrix for this state is thus\,:
\begin{eqnarray}
 M^{(1)}(|\psi^{(a)} (\omega) \rangle) &=& 
\left( \begin{array}{cc}
1 & S^2 \\
S^2 & S^2(2S^2 + C^2)
\end{array} \right) , \nonumber \\
 {\rm det} M^{(1)}(|\psi^{(a)} (\omega) \rangle) &=& S^2(S^2 + C^2) \geq 0,
\label{m4.7}
\end{eqnarray}
where $S$ and $C$ are given in Eq. (\ref{m4.6}). Thus these states have
super-PS, and the QO-nonclassicality does not show up, or is missed, at the
Mandel level.
\subsection{Mandel matrices for two-mode states}
The two-mode generalization of the Mandel matrix idea leads naturally to a more
intricate classification of states. We consider only states
$\hat{\rho}^{(ab)}_{D}$ conserving, i.e., commuting with, $\hat{N}_a +
\hat{N}_b$. First we develop the analogue of the positivity property
(\ref{m4.1}), the statement that the number fluctuation $(\Delta N_a)^2$ for a
PND $\{p(n) \}$ can never be negative. Define a column and row vector with
number conserving operator entries as follows
\begin{eqnarray}
\hat{C} &=& 
\left( \begin{array}{c}
\hat{a}^{\dagger}\\
\hat{b}^{\dagger}
\end{array} \right)
\otimes
\left( \begin{array}{c}
\hat{a}^{}\\
\hat{b}^{}
\end{array} \right)
=
\left( \begin{array}{c}
\hat{N}_a \\
\hat{a}^{\dagger} \hat{b}\\
\hat{b}^{\dagger} \hat{a} \\
\hat{N}_b
\end{array} \right), \nonumber \\
\hat{C}^{\dagger} &=&
\left( \begin{array}{cccc}
\hat{N}_a &
\hat{b}^{\dagger} \hat{a}&
\hat{a}^{\dagger} \hat{b} &
\hat{N}_b
\end{array} \right).
\label{m4.8}
\end{eqnarray}
With their help next define a $5\times 5$ matrix with operator entries 
and which is `hermitian' like $\hat{N}$ in Eq. (\ref{m3.6}), and also
`positive definite'\,:
\begin{eqnarray}
\hat{\Sigma}&=&
\left(\begin{array}{c}
1\\
\hat{C}
\end{array}
\right)
\begin{array}{c}
\left(
\begin{array}{cc}
1 & \hat{C}^{\dagger}
\end{array}\right)
\\
\\
\end{array}
=
\left( \begin{array}{cc}
1 & \hat{C}^{\dagger} \\
\hat{C} & \hat{C}\hat{C}^{\dagger}
\end{array} \right) \nonumber \\
 &=&
\left( \begin{array}{ccccc}
1& \hat{N}_a & \hat{b}^{\dagger}\hat{a}& \hat{a}^{\dagger} \hat{b} &\hat{N}_b \\
\hat{N}_a &\hat{N}_{a}^{2} & \hat{N}_a \hat{b}^{\dagger}\hat{a}&\hat{N}_a 
\hat{a}^{\dagger} \hat{b} &\hat{N}_a \hat{N}_b \\
\hat{a}^{\dagger} \hat{b} & \hat{a}^{\dagger} \hat{b}\hat{N}_a  & 
\hat{a}^{\dagger} \hat{b}\hat{b}^{\dagger}\hat{a} & (\hat{a}^{\dagger} \hat{b})^2
&\hat{a}^{\dagger} \hat{b}\hat{N}_b \\
\hat{b}^{\dagger}\hat{a} &\hat{b}^{\dagger}\hat{a}\hat{N}_a  & 
(\hat{b}^{\dagger}\hat{a})^2  & \hat{b}^{\dagger}\hat{a}\hat{a}^{\dagger} 
\hat{b}  & \hat{b}^{\dagger}\hat{a}\hat{N}_b \\
\hat{N}_b & \hat{N}_b \hat{N}_a & \hat{N}_b\hat{b}^{\dagger}\hat{a} & 
\hat{N}_b\hat{a}^{\dagger} \hat{b} & \hat{N}_{b}^{2}
\end{array} \right).
\label{m4.9}
\end{eqnarray}
Given a state $\hat{\rho}^{(ab)}_{D}$, by taking entrywise expectation values of
the operators in $\hat{\Sigma}$ we get the $5 \times 5$ numerical hermitian
matrix $\Sigma$ which is clearly hermitian positive semidefinite and generalizes
(\ref{m4.1})\,:
\begin{eqnarray}
\Sigma= \langle \hat{\Sigma} \rangle &=&
{\rm Tr} (\hat{\rho}^{(ab)}_{D}
\left( \begin{array}{cc}
1 & \hat{C}^{\dagger} \\
\hat{C} & \hat{C}\hat{C}^{\dagger}
\end{array} \right)) \nonumber \\
&=& \left( \begin{array}{cc}
1 & \langle \hat{C}^{\dagger} \rangle \\
\langle \hat{C} \rangle & \langle \hat{C}\hat{C}^{\dagger} \rangle
\end{array} \right) \geq 0\,.
\label{m4.10}
\end{eqnarray}
We now define the {\em two-mode Mandel matrix} for the state
$\hat{\rho}^{(ab)}_{D}$ by replacing $\hat{C}\hat{C}^{\dagger}$ in Eq.
(\ref{m4.10}) by its normal ordered expression (the entries in $\hat{C}$ and
$\hat{C}^{\dagger}$ are already in the normal ordered form) \cite{arvind98}\,:
\begin{eqnarray} 
\hat{B} &=& :\hat{C}\hat{C}^{\dagger}:, \nonumber \\ 
M^{(2)}(\hat{\rho}^{(ab)}_{D}) &=& 
{\rm Tr} (\hat{\rho}^{(ab)}_{D}
\left( \begin{array}{cc}
1 & \hat{C}^{\dagger} \\
\hat{C} & \hat{B}
\end{array} \right)) \nonumber \\
& = & \left( \begin{array}{ccccc}
1 & \langle {\hat{a}}^{\dagger}{\hat{a}} \rangle & \langle {\hat{b}}^{\dagger} \hat{a} \rangle 
& \langle {\hat{a}}^{\dagger}{\hat{b}} \rangle & \langle {\hat{b}}^{\dagger}{\hat{b}} \rangle \\
\langle {\hat{a}}^{\dagger} {\hat{a}} \rangle & 
\langle {\hat{a}}^{\dagger 2} {\hat{a}}^{2} \rangle 
& \langle {\hat{a}}^{\dagger}{\hat{b}}^{\dagger} {\hat{a}}^{2} \rangle & 
\langle {\hat{a}}^{\dagger 2} {\hat{a}} {\hat{b}} \rangle & 
\langle {\hat{a}}^{\dagger}{\hat{b}}^{\dagger}  {\hat{a}} {\hat{b}}\rangle \\
\langle {\hat{a}}^{\dagger}{\hat{b}}\rangle & 
\langle {\hat{a}}^{\dagger 2} {\hat{a}} {\hat{b}} \rangle & 
\langle {\hat{a}}^{\dagger}{\hat{b}}^{\dagger}  {\hat{a}} {\hat{b}}\rangle & 
\langle {\hat{a}}^{\dagger 2} {\hat{b}}^{2} \rangle &
\langle {\hat{a}}^{\dagger}{\hat{b}}^{\dagger}{{\hat{b}}}^{2} \rangle \\
\langle {\hat{b}}^{\dagger}{\hat{a}} \rangle & 
\langle {\hat{a}}^{\dagger}{\hat{b}}^{\dagger} {{\hat{a}}}^{2} \rangle & 
\langle {\hat{b}}^{\dagger 2}{\hat{a}}^{2} \rangle & 
\langle {\hat{a}}^{\dagger}{\hat{b}}^{\dagger}  {\hat{a}}{\hat{b}}\rangle & 
\langle {\hat{b}}^{\dagger 2}{\hat{a}} {\hat{b}} \rangle \\ 
\langle {\hat{b}}^{\dagger}{\hat{b}} \rangle & 
\langle {\hat{a}}^{\dagger}{\hat{b}}^{\dagger}  {\hat{a}}{\hat{b}}\rangle & 
\langle {\hat{b}}^{\dagger 2}{\hat{a}} {\hat{b}} \rangle & 
\langle {\hat{a}}^{\dagger} {{\hat{b}}^{\dagger}} {{\hat{b}}}^{2}  \rangle & 
\langle {\hat{b}}^{\dagger 2} {\hat{b}}^{2} \rangle 
\end{array} \right) \nonumber \\ 
& = & \int_{0}^{\infty} dI_a \int_{0}^{\infty} dI_b \int_{0}^{2 \pi} \frac{d \theta}{2 \pi} 
P(I_a, I_b, \theta) \chi(I_a , I_b , \theta ) \, {\chi(I_a , I_b , \theta)}^{\dagger}\nonumber \\
\chi(I_a , I_b , \theta )&=&\left( \begin{array}{ccccc} 
1 & I_a & \sqrt{I_a I_b} e^{+i \theta} &\sqrt{I_a I_b} e^{-i \theta} & I_b  
\end{array} \right)^{T} 
\label{m4.11}
\end{eqnarray}
The superscript ${}^{(2)}$ indicates that we are dealing with a two-mode state,
and this Mandel matrix is $5 \times 5$ hermitian but not necessarily positive
semidefinite. 

Before presenting a classification of two-mode states based on the $5 \times 5$
Mandel matrix, we introduce a useful derived object. This is the $2 \times 2$
Mandel matrix associated with a general single mode defined as a linear
combination of the modes $a$ and $b$, the reduced subsystem state of this chosen
single mode being calculated with respect to the two-mode state
$\hat{\rho}_{D}^{(ab)}$. The definition of the annihilation operator $\hat{A}$
of such a mode and then of its Mandel matrix are\,:
\begin{eqnarray}
\hat{A} = \alpha \hat{a} + \beta \hat{b},&& \,\,\, |\alpha|^2 + 
|\beta|^2 =1:\,\,\,\,\,[\hat{A} \hat{A}^{\dagger}]={1\!\!1} \nonumber \\
M^{(2,1)} (\hat{\rho}^{(ab)}_{D};\alpha, \beta) &=& 
\left( \begin{array}{cc}
1 & \langle  \hat{A}^{\dagger} \hat{A} \rangle \\
\langle  \hat{A}^{\dagger} \hat{A} \rangle &
\langle \hat{A}^{\dagger 2} \hat{A}^{2} \rangle
\end{array} \right) \nonumber \\
&=& Y(\alpha, \beta)^{\dagger} M^{(2)} ({\hat{\rho}}^{(ab)}_{D}) Y(\alpha, 
\beta), \nonumber \\
 Y(\alpha, \beta) &=&
\left( \begin{array}{cc}
1 & 0 \\
0 & \\
0 &  {\psi}_{0}(\alpha, \beta) \\
0 & \\
0& 
\end{array} \right), \nonumber \\
 {\psi}_{0}(\alpha, \beta) &=& 
\left( \begin{array}{c}
\alpha \\
\beta
\end{array} \right)
\otimes 
\left( \begin{array}{c}
{\alpha}^{*} \\
{\beta}^{*}
\end{array} \right)
=
\left( \begin{array}{c}
\alpha {\alpha}^{*} \\
\alpha {\beta}^{*} \\
\beta {\alpha}^{*}\\
\beta {\beta}^{*}
\end{array} \right).
\label{m4.12}  
\end{eqnarray} 
The dependence of $\hat{A}$ on $\alpha, \beta$ is left implicit. The superscript
$(2,1)$ at the start of the above equations indicates that we are dealing with a
general single-mode Mandel matrix obtained from the two-mode Mandel matrix for
the $a-b$ system in the state $\hat{\rho}^{(ab)}_{D}$, by focussing on a
variable linear combination $\hat{A}$ of $\hat{a}$ and $\hat{b}$. As we will
immediately see, for two-mode states both the $5 \times 5$  matrix
$M^{(2)}(\hat{\rho}^{(ab)}_{D})$ and the $2 \times 2$  matrix
$M^{(2,1)}(\hat{\rho}^{(ab)}_{D}; \alpha,\beta)$ are important.

The two-mode definitions of Mandel-type nonclassicality, sub-Poissonian
statistics (sub-PS), super-Poissonian statistics (super-PS), etc are now as
follows\,:
\begin{eqnarray}
\{ \hat{\rho}_{D}^{(ab)} \,\,\textrm{is QO-cl}\,\,& \Leftrightarrow& \,\,
P(I_a , I_b, \theta) \geq 0 \}\nonumber \\
& \Rightarrow &\{ M^{(2)}(\hat{\rho}^{(ab)}_{D}) \geq 0 \,\, 
\Leftrightarrow  
\,\,\hat{\rho}_{D}^{(ab)}
\textrm{has super-PS}\}; \nonumber \\
M^{(2)}(\hat{\rho}^{(ab)}_{D}) \not\geq 0 \,\, 
&\Leftrightarrow& \,\,\{ \hat{\rho}_{D}^{(ab)}
\textrm{is QO-noncl}, \,\, \textrm{has sub-PS}\}.
\label{m4.13}
\end{eqnarray}
In the definition of super-PS, we used Eq. (\ref{m4.11}). The sub-PS case can be
usefully separated into two types, depending on whether or not the nonpositivity
of the $5\times 5$ matrix $M^{(2)}(\hat{\rho}^{(ab)}_{D})$ is visible already at
the single-mode level for some choice of coefficients $\alpha$, $\beta$. Thus we
define\,:
\begin{eqnarray}
\hat{\rho}^{(ab)}_{D} \,\,\textrm{has Type I sub-PS} \,\, 
&\Leftrightarrow& \,\,
M^{(2,1)}(\hat{\rho}^{(ab)}_{D}; \alpha, \beta) \not\geq 0 \,\, \textrm{for some}
\,\, \alpha, \,\, \beta; \nonumber \\
\hat{\rho}^{(ab)}_{D} \,\,\textrm{has Type II sub-PS} \,\, 
&\Leftrightarrow& \,\,M^{(2)}(\hat{\rho}^{(ab)}_{D}) \not\geq 0,
\,\,{\rm but} \nonumber \\
&& M^{(2,1)}(\hat{\rho}^{(ab)}_{D}; \alpha, \beta) \geq 0 \,\, \textrm{for all}
\,\, \alpha, \,\, \beta.
\label{m4.14}
\end{eqnarray} 
The physical meaning is that in Type I sub-PS, the Mandel level of QO
nonclassicality is easy to detect already in terms of a suitable single-mode
combination; while in Type II sub-PS, such nonclassicality is hidden or
intrinsically two-mode in character \cite{arvind98}.

For calculational purposes one can pass from the $5\times 5$ Mandel matrix
$M^{(2)}(\hat{\rho}^{(ab)}_{D})$ to a slightly simpler $4\times 4$ matrix as follows. From Eq. (\ref{m4.11}),

\begin{eqnarray}
M^{(2)}(\hat{\rho}^{(ab)}_{D})& = & {\rm Tr} (\hat{\rho}^{(ab)}_{D}
\left( \begin{array}{cc}
1 & \hat{C}^{\dagger} \\
\hat{C} & \hat{B}
\end{array} \right)) \,\, = \,\, 
\left( \begin{array}{cc}
1 & {C}^{\dagger} \\
{C} & {B}
\end{array} \right)\, , \nonumber \\
 C &=& {\rm Tr} (\hat{\rho}^{(ab)}_{D}\hat{C}), \,\, B = {\rm Tr}(\hat{\rho}^{(ab)}_{D}\hat{B}).
\label{m4.15}
\end{eqnarray}
(When necessary the state will be indicated as argument of $C, B$). Then it is easy to see
that
\begin{eqnarray}
&&M^{(2)}(\hat{\rho}^{(ab)}_{D}) \geq 0 \,\, \Leftrightarrow \,\, \Gamma = B - C C^{\dagger}
\geq 0 \, , \nonumber \\
&&M^{(2)}(\hat{\rho}^{(ab)}_{D}) \not\geq 0 \,\, \Leftrightarrow \,\, \Gamma \not\geq 0. 
\label{m4.16}
\end{eqnarray} 
Thus the $4 \times 4$ hermitian matrix $\Gamma$ determines whether we have
super-PS or sub-PS. For the separation of the latter into Type I and Type II, we
have for any complex 2-vector $\phi = \left( \begin{array}{c} \phi_{1} \\
\phi_{2} \end{array} \right)$\,: 





\begin{eqnarray}
&&{\phi}^{\dagger} M^{(2,1)}(\hat{\rho}^{(ab)}_{D};\alpha, \beta) \phi = \nonumber \\
&&|\phi_{1} + \phi_{2} C^{\dagger} \psi_{0} (\alpha, \beta) |^2
+ |\phi_{2}|^2 \psi_{0} (\alpha, \beta)^{\dagger} \Gamma \psi_{0} (\alpha, \beta).
\label{m4.17}
\end{eqnarray}
So given $M^{(2)}(\hat{\rho}^{(ab)}_{D}) \not\geq 0$, we are able to say\,:
\begin{eqnarray}
\textrm{Type I sub-PS} \,\, & \Leftrightarrow & \,\,\psi_{0} (\alpha, \beta)^{\dagger} 
\Gamma \psi_{0} (\alpha, \beta) < 0 \,\,{\rm for \,\, some} \,\, \alpha, \,\, \beta;
\nonumber \\
\textrm{Type II sub-PS} \,\,& \Leftrightarrow & \,\,\psi_{0} (\alpha, \beta)^{\dagger} 
\Gamma \psi_{0} (\alpha, \beta) \geq 0 \,\,{\rm for \,\, all} \,\, \alpha, \,\, \beta. 
\label{m4.18}
\end{eqnarray}
Indeed we easily find from Eqs. (\ref{m4.12}, \ref{m4.16}) that
\begin{eqnarray}
&&{\rm det}\,M^{(2,1)}(\hat{\rho}^{(ab)};\alpha, \beta) = 
{\psi_{0} (\alpha, \beta)}^{\dagger} \Gamma \psi_{0} (\alpha, \beta).
\label{m4.19}
\end{eqnarray}

\subsection{Examples of two-mode states and Mandel matrices}
We consider two examples. We have seen in Section II that a single-mode QO-noncl
state, when combined with a second  ancilla mode in vacuum (or in a coherent
state) and then passed through a nontrivial $U(2)$ BS, always results at the
output in a two-mode state exhibiting NPT entanglement. We study  this as the first example.

The two-mode state in question is given in Eq. (\ref{m2.16}). It is
understandable that its $5 \times 5$ Mandel matrix is obtainable from the $2
\times 2$ Mandel matrix associated with the single mode input state
$\hat{\rho}^{(a)}_{D}$. Straightforward calculation shows that\,:
\begin{eqnarray}
\hat{\rho}_{D}^{(ab)} &=& \hat{U}(u) \{ \hat{\rho}_{D}^{(a)} \otimes|0 \rangle_b {}_b\langle 0 |\}
\hat{U}(u)^{-1} , \,\,\,\, u \in U(2): \nonumber \\
M^{(2)}(\hat{\rho}_{D}^{(ab)}) &=& W(u)^{\dagger}M^{(1)}(\hat{\rho}_{D}^{(a)}) W(u) , \nonumber \\
W(u)& = & 
\left( \begin{array}{ccccc}
1 & 0& 0& 0& 0 \\
0 & u_{11}^{*}u_{11} &u_{21}^{*}u_{11} &u_{11}^{*}u_{21} &u_{21}^{*}u_{21} 
\end{array} \right), \nonumber \\
W(u) W(u)^{\dagger} &=& {1\!\!1}_{2 \times 2}.
\label{m4.20} 
\end{eqnarray}
Next using (\ref{m4.12}) we can immediately obtain the variable single-mode projection of this
two-mode Mandel matrix\,:
\begin{eqnarray}
M^{(2,1)} (\hat{\rho}_{D}^{(ab)}; \alpha, \beta) & = & Y( \alpha, \beta)^{\dagger} 
W(u)^{\dagger} M^{(1)}(\hat{\rho}_{D}^{(a)}) W(u)Y( \alpha, \beta)  \nonumber \\
& = & \left( \begin{array}{cc}
1 & 0 \\
0 & |\xi|^2
\end{array} \right)
M^{(1)}(\hat{\rho}_{D}^{(a)})
\left( \begin{array}{cc}
1 & 0 \\
0 & |\xi|^2
\end{array} \right), \nonumber \\
\xi &=& u_{11} \alpha + u_{22} \beta.
\label{m4.21}
\end{eqnarray}
From these expressions and the results of Section II, we find that the two-mode
states produced from single-mode states in the above manner have the following
significant properties\,:
\begin{eqnarray}
&&(i)\,\, \hat{\rho}^{(a)}_{D} \,\,\textrm{has QO-noncl PND} \,\, \Rightarrow \,\,
\hat{\rho}^{(ab)}_{D} \,\,\textrm{has NPT entanglement}; \nonumber \\
&&(ii)\,\,\hat{\rho}^{(a)}_{D} \,\,\textrm{ has super-PS}\,\, \Rightarrow \,\,
\hat{\rho}^{(ab)}_{D} \,\, \textrm{has super-PS};\nonumber \\
&&(iii)\,\, \hat{\rho}^{(a)}_{D} \,\,\textrm{ has sub-PS}\,\, \Rightarrow \,\,
\hat{\rho}^{(ab)}_{D} \,\,
\textrm{has Type I sub-PS},\,\, \nonumber \\
&&\hspace{4cm}M^{(2,1)} ({\rho}_{D}^{(ab)}; \alpha, \beta)
\not\geq 0 \,\,{\rm for \,\, every} \,\, \alpha, \beta. 
\label{m4.22}
\end{eqnarray}
Of course only properties $(ii)$ and $(iii)$ involve the Mandel matrix analysis;
it is significant that in $(iii)$, every single-mode combination of the modes
$a$ and $b$ displays sub-PS. To this we can add the following\,: states
$\hat{\rho}_{D}^{(ab)}$ obtained from states $\hat{\rho}_{D}^{(a)}$ via Eq.
(\ref{m4.20}) can never display Type-II sub-PS; and any sub-PS in
$\hat{\rho}_{D}^{(a)}$ leads to both Type-I sub-PS and NPT entanglement in
$\hat{\rho}_{D}^{(ab)}$.

The second example is the two-mode generalization of the squeezed vacuum state
defined for a single mode in Eq. (\ref{m4.5}). We take independent complex
$\xi$, $\xi'$ or $\omega$, $\omega'$ and define\,:
\begin{eqnarray}
&&|\psi^{(ab)} (\omega, \omega') \rangle = |\psi^{(a)} (\omega) \rangle \otimes
|\psi^{(b)} (\omega') \rangle,
\label{m4.23}
\end{eqnarray}
with the second factor involving an exponential in $\hat{b}^{\dagger 2}$ and
$\hat{b}^2$ applied to $| 0 \rangle_b$. This pure state is clearly also QO-noncl, 
but it is manifestly a product state of Schmidt rank one. Unlike 
the single mode case in Eq. (\ref{m4.7}), however, now
the QO nonclassicality shows up at the Mandel level. The $5 \times 5$ Mandel
matrix for the state (\ref{m4.23}) is easily found using Eqs. (\ref{m4.6}) and 
their $b$-mode analogues\,:
\begin{eqnarray}
&&M^{(2)}(|\psi^{(ab)} (\omega, \omega') \rangle) = 
\left( \begin{array}{cc}
1 & C^{\dagger} \\
C & B
\end{array} \right) , \nonumber \\
&&C^{\dagger} = \left( \begin{array}{cccc}
 S^2&0&0&S'^2
\end{array} \right) , \nonumber \\
&&B = \left( \begin{array}{cccc}
S^2(2S^2+C^2) &0&0&S^2 S'\,^2 \\
0 & S^2 S'\,^2 & e^{i \eta}SCS'C' &0 \\
0 & e^{-i \eta}SCS'C' & S^2 S'\,^2 &0 \\
S^2 S'\,^2&0&0& S'\,^2(2S'\,^2+ C'\,^2) 
\end{array} \right) , \nonumber \\
&&\eta = {\rm arg}\xi' \xi^*
\label{m4.24}
\end{eqnarray}
Here $S'$ and $C'$ are defined as in Eq. (\ref{m4.6}) but in terms of $\xi'$. The $4 \times 4$
matrix $\Gamma$ of Eq. (\ref{m4.16}) is\,:
\begin{eqnarray}
\Gamma = \left( \begin{array}{cccc}
S^2(S^2+C^2) &0&0&0 \\
0 & S^2 S'\,^2 & e^{i \eta}SCS'C' &0 \\
0 & e^{-i \eta}SCS'C' & S^2 S\,'^2 &0 \\
0&0&0& S'\,^2(S'\,^2+ C'\,^2)
\end{array} \right).
\label{m4.25}
\end{eqnarray}
The eigenvalues of $\Gamma$ are $S^2(S^2+C^2)$, $S'^2(S'^2+ C'^2)$, $SS'(SS' + CC')$
and $SS'(SS'- CC')$. Assuming that $\xi$, $\xi'$ are both non zero, the last
eigenvalue is negative, leading by Eq. (\ref{m4.16}) to the conclusion that
$M^{(2)}(|\psi^{(ab)} (\omega ,\omega') \rangle) \not\geq 0$ or that the state
$|\psi^{(ab)} (\omega ,\omega') \rangle$ has sub-PS. This is an interesting and
somewhat nonintuitive result since we have seen in Eq. (\ref{m4.7}) that each
factor in the product state $|\psi^{(ab)} (\omega, \omega') \rangle$ has
super-PS. We must now see whether it is Type I or Type II. For this we must
compute the `expectation value' of $\Gamma$ in Eq. (\ref{m4.25}) for the
four-component column vector $\psi_0 (\alpha, \beta)$ as required by Eq.
(\ref{m4.18})\,:
\begin{eqnarray}
&&\!\!\!\!\!\!\!\!\!\psi_{0} (\alpha, \beta)^{\dagger} \Gamma \psi_{0} (\alpha, \beta) = \nonumber \\
&&( |\alpha|^2 S^2 + |\beta|^2 S'^2)^2 + |\alpha|^4 S^2 C^2 
 + |\beta|^4 S'^2 C'^2 + 2SCS'C' \Re{(e^{i \eta}(\alpha^* \beta)^2)} \nonumber \\
&& \,\,\,\,\,\,\,\, \geq ( |\alpha|^2 S^2 + |\beta|^2 S'^2)^2 + (|\alpha|^2SC-|\beta|^2S'C')^2 > 0,
\label{m4.26}
\end{eqnarray}  
since $\Re{(e^{i \eta}(\alpha^* \beta)^2)} \geq -|\alpha|^2|\beta|^2$. It
follows that the sub-PS of the product state $|\psi^{(ab)} (\omega, \omega')
\rangle$ is of Type II, it is hidden or intrinsic. This is consistent with the
fact that the individual factors $|\psi^{(a)} (\omega) \rangle$ and $|\psi^{(b)}
(\omega') \rangle$ are both super-PS.

\section{Two-mode Mandel level nonclassicality to entanglement by BS action}We
now consider a  two-mode state $\hat{\rho}_{D}^{(ab)}$ which is QO nonclassical
and of such a nature that this property is seen at the level of the Mandel
matrix, i.e., $M^{(2)}(\hat{\rho}_{D}^{(ab)}) \not\geq 0$. In such a case, even
if $\hat{\rho}_{D}^{(ab)}$ is of product or separable form, its passage through
a $U(2)$ BS could result in an entangled state, possibly of NPT type. Our aim is
now to see quantitatively how much beyond the nonpositivity of the Mandel matrix
one has to go to reach NPT entanglement. We set up the general framework to
examine this, then illustrate it by an example. For simplicity we use a 50:50 BS
rather than one corresponding to a general $u$ $\in$ $U(2)$.

 We choose the
$U(2)$ element and corresponding unitary operator action as follows
\cite{campos89, arvind96}\,: 
\begin{equation}
u_0 = \frac{1}{\sqrt{2}}
\left( \begin{array}{cc}
1 & 1 \\
-1 & 1
\end{array} \right) \in U(2) : \,\,
\hat{U}_{0}^{-1}
\left( \begin{array}{cc}
\hat{a} & \hat{a}^{\dagger} \\
\hat{b} & \hat{b}^{\dagger}
\end{array} \right) \hat{U}_0 \,\,
= u_0
\left( \begin{array}{cc}
\hat{a} & \hat{a}^{\dagger} \\
\hat{b} & \hat{b}^{\dagger}
\end{array} \right).
\label{m5.1}
\end{equation}
At the operator level, action by conjugation on $\hat{C}$, $\hat{C}^{\dagger}$,
$\hat{B}$ of Eqs. (\ref{m4.8}, \ref{m4.11}) is\,:
\begin{eqnarray}
\hat{U}_{0}^{-1} \hat{C} \hat{U}_{0} &=& V_0 \hat{C}, \,\,
\hat{U}_{0}^{-1} \hat{C}^{\dagger} \hat{U}_0 = 
\hat{C}^{\dagger} V_{0}^{T}
, \,\,\hat{U}_{0}^{-1} \hat{B} \hat{U}_{0} = V_0 \hat{B} V_{0}^{T}, \nonumber \\
V_0 &=& u_0 \otimes u_0 =
\frac{1}{2}
\left( \begin{array}{cccc}
1 & 1 & 1 & 1 \\
-1 & 1 & -1 &1 \\
-1 &-1 & 1 & 1 \\
1 &-1 & -1 & 1 
\end{array} \right).
\label{m5.2}
\end{eqnarray}
Then when the state $\hat{\rho}^{(ab)}_{D}$ is transformed by this BS action to
\begin{equation}
\hat{\rho}'\,^{(ab)}_{D} = \hat{U}_0 \hat{\rho}^{(ab)}_{D} \hat{U}_0^{-1},
\label{m5.3}
\end{equation}
the new Mandel matrix is given by a transformation using $V_{0}$\,: 
\begin{eqnarray}
M^{(2)} (\hat{\rho}^{(ab)}_{D}) &=& 
\left( \begin{array}{cc}
1 & C^{\dagger} \\
C & B
\end{array} \right) \rightarrow \nonumber \\
M^{(2)} (\hat{\rho}'\,^{(ab)}_{D}) &=&
\left( \begin{array}{cc}
1 & C'^{\dagger} \\
C' & B'
\end{array} \right) = 
{\rm Tr}( \hat{\rho}'\,^{(ab)}_{D} 
\left( \begin{array}{cc}
1 & \hat{C}^{\dagger} \\
\hat{C} & \hat{B}
\end{array} \right)) \nonumber \\
&=& 
\left( \begin{array}{cc}
1 & 0 \\
0 & V_0
\end{array} \right)
\left( \begin{array}{cc}
1 & C^{\dagger} \\
C & B
\end{array} \right)
\left( \begin{array}{cc}
1 & 0 \\
0 & V_{0}^{T}
\end{array} \right), \nonumber \\
C' &=& V_0 C, \,\, B' = V_0 B V_{0}^{T}.
\label{m5.4}
\end{eqnarray}
Thus $\Gamma'$ is related to $\Gamma$ by {\em congruence}\,:
\begin{equation}
\Gamma' = B' - C' C'^{\dagger} = V_0 \Gamma V_{0}^{T}.
\label{m5.5}
\end{equation}

To test next whether $\hat{\rho}'\,^{(ab)}_{D}$ is NPT entangled, we pass to its
partial transpose $\hat{\rho}'\,^{(ab)PT}_{D}$ and evaluate the `expectation
value' of a suitably chosen nonnegative hermitian operator with respect to it.
If this turns out to be negative, then  the output state 
$\hat{\rho}'\,^{(ab)}_{D}$ is definitely NPT entangled. To construct such a test
which involves as closely as possible the use of
$M^{(2)}(\hat{\rho}'\,^{(ab)}_{D})$, hence of $M^{(2)}(\hat{\rho}^{(ab)}_{D})$,
plus something additional, we should use a `matrix of operators' similar in
structure to $\left(\begin{array}{c} 1 \\ \hat{C}\end{array}\right)
\begin{array}{c} \left(1\,\,\hat{C}^{\dagger}\right) \\ \\ \end{array}$, i.e,
making up a `hermitian nonnegative' matrix of operator entries, such that when
the partial transpose operation is switched from $\hat{\rho}'\,^{(ab)PT}_{D}$ to
this `matrix', we obtain the expectation values of $\hat{C}$,
$\hat{C}^{\dagger}$ and $\hat{B}$ in $\hat{\rho}'\,^{(ab)}_{D}$, plus something
additional. Now we have seen in the passage from Eq. (\ref{m3.7}) to Eq.
(\ref{m3.8}) that the PT operation converts $\hat{b}^{\dagger j} \hat{b}^{k}$ to
$\hat{b}^{\dagger k} \hat{b}^{j}$, and $\hat{b}^{j } \hat{b}^{\dagger k}$ to
$\hat{b}^{k } \hat{b}^{\dagger j}$. Keeping these motivations and facts in mind
we construct a $5 \times 5$ matrix of operators as follows\,:

\begin{eqnarray}
\hat{E} =
\left( \begin{array}{c}
\hat{a}^{\dagger} \hat{a} \\
\hat{a}^{\dagger}\hat{b}^{\dagger} \\
\hat{a}^{} \hat{b} \\
\hat{b}^{\dagger} \hat{b}
\end{array} \right) , \,\,
\hat{E}^{\dagger} &=& 
\left( \begin{array}{cccc}
\hat{a}^{\dagger} \hat{a} &
\hat{a}^{} \hat{b} &
\hat{a}^{\dagger}\hat{b}^{\dagger} &
\hat{b}^{\dagger} \hat{b}
\end{array} \right)\,\, \rightarrow \nonumber \\
\left\{ \left( \begin{array}{c}
1 \\
\hat{E}
\end{array} \right)
\begin{array}{c}
\left( \begin{array}{cc}
1 & \hat{E}^{\dagger}
\end{array} \right)
\\ \\
\end{array} \right\}^{PT} &=&
\left( \begin{array}{cc}
1 & \hat{C}^{\dagger} \\
\hat{C} & \hat{B}
\end{array} \right)
+ \left( \begin{array}{cc}
0 & 0 \\
0 & \hat{Y}
\end{array} \right), \nonumber \\
\hat{Y} &=&
\left( \begin{array}{ccccccc}
\hat{a}^{\dagger} \hat{a} && 0 && \hat{a}^{\dagger} \hat{b} && 0\\
0 && 0 && 0 && 0  \\
\hat{b}^{\dagger} \hat{a} && 0 && \hat{a}^{\dagger} \hat{a}+ \hat{b}^{\dagger} \hat{b} +1 &&
\hat{b}^{\dagger} \hat{a} \\
0 && 0 && \hat{a}^{\dagger} \hat{b} && \hat{b}^{\dagger} \hat{b}
\end{array} \right) .
\label{m5.6}
\end{eqnarray}
We see that in the process of expressing the various operators involved in
normal ordered form, as anticipated an additional piece $\hat{Y}$ linear in the
entries of $\hat{C}$ appears. Then a test for NPT entanglement of
$\hat{\rho}'\,_{D}^{(ab)}$ is to evaluate 

\begin{eqnarray}
{\rm Tr}(\hat{\rho}'\,^{(ab)PT}_{D} \begin{array}{c}
\\
\left( \begin{array}{c}
1 \\
\hat{E}
\end{array} \right) \end{array}
\left( \begin{array}{cc}
1 & \hat{E}^{\dagger}
\end{array} \right)) \,\, &=& \,\,
{\rm Tr}(\hat{\rho}'\,^{(ab)}_{D}
\begin{array}{c}
\\
\left\{
\left( \begin{array}{c}
1 \\
\hat{E}
\end{array}\right)
\begin{array}{c}
\left( \begin{array}{cc}
1 & \hat{E}^{\dagger}
\end{array} \right)
\\ \\
\end{array}
\right\}^{PT}
\end{array}) \nonumber \\
&=& M^{(2)}(\hat{\rho}'\,^{(ab)}_{D}) + 
%
%
\left(\begin{array}{ccc} \begin{array}{c} 0 \end{array} & \begin{array}{c}
\vdots \end{array} & \begin{array}{ccccccc} 0&&0&&0&&0 \\ \end{array} \\
\begin{array}{c} \cdots \end{array} && \begin{array}{cccc} \cdots & \cdots &
\cdots & \cdots \end{array} \\ \begin{array}{c} 0 \\ 0 \\ 0\\ 0 \end{array} &
\begin{array}{c} \vdots \\ \vdots \\ \vdots \\ \vdots \end{array} & Y'
\end{array} \right), \nonumber \\
%
%
%
%
%
%
%
%
Y' &=& \left( \begin{array}{ccccccc} C'_1 && 0 && C'_2 && 0 \\ 0 && 0 && 0 && 0
\\ C'_3 && 0 && C'_1 + C'_4 +1 && C'_3 \\ 0 && 0 && C'_2 && C'_4 \end{array}
\right) , \label{m5.7} 
\end{eqnarray}
and see if this matrix is indefinite. By Eq. (\ref{m5.4}), the complete $5
\times 5$ matrix here is a congruence transformation applied to the initial
state Mandel matrix $M^{(2)}(\hat{\rho}^{(ab)}_{D})$ plus a $4 \times 4$ piece
coming from $\hat{Y}$, namely it is\,:
\begin{eqnarray}
\left( \begin{array}{cc}
1 & 0\\
0 & V_0
\end{array} \right) 
\left( \begin{array}{cc}
1 & C^{\dagger}\\
C & B
\end{array} \right)
\left( \begin{array}{cc}
1 & 0\\
0 & V_{0}^{T}
\end{array} \right) + 
\left( \begin{array}{cc}
0 & 0\\
0 & Y'
\end{array} \right)
 = 
\left( \begin{array}{cc}
1 & C^{\dagger}V_{0}^{T}\\
V_0 C & V_0 B V_{0}^{T} + Y'
\end{array} \right).
\label{m5.8}
\end{eqnarray}
Therefore by Eq. (\ref{m4.16}) the positivity or otherwise of the matrix (\ref{m5.7})
is equivalent to the positivity or otherwise of either of the two following
$4 \times 4$ matrices at the level of $\Gamma$\,:
\begin{eqnarray}
\Omega & = & \Gamma + V_{0}^{T} Y' V_{0}\, ,\nonumber \\
V_0 \Omega V_{0}^{T} & = & V_0 \Gamma V_{0}^{T} + Y'. 
\label{m5.9}
\end{eqnarray}
Nonpositivity of either $\Omega$ or $V_0 \Omega V_{0}^{T}$ is proof of NPT
entanglement of the output state  $\hat{\rho}'\,^{(ab)}_{D}$. It is interesting
to see the precise quantitative manner in which this test goes beyond
examination of $M^{(2)}(\hat{\rho}^{'(ab)}_{D})$ or $\Gamma'$ alone.

\subsection{An illustrative example}
To see how the general procedure developed above works, we study a
family of states which is analytically quite simple. We begin with the family of two-mode
pure states of infinite Schmidt rank,
 \begin{equation}
| \mu \rangle = e^{- \frac{1}{2} |\mu|^2} \sum_{n=0}^{\infty} \frac{\mu^{n}}{\sqrt{n!}}
| n, n \rangle , \,\, \mu \in \cal{C} ,
\label{m5.10}
\end{equation}
form the density matrix $\hat{\rho}^{(ab)} = | \mu \rangle \langle \mu |$, and
pass to $\hat{\rho}^{(ab)}_{D}$ via Eq. (\ref{m3.3})\,:
\begin{equation}
\hat{\rho}^{(ab)}_{D} = e^{- \lambda} \sum_{n=0}^{\infty} \frac{\lambda^n}{n!}
| n, n \rangle \langle n, n | , \,\, \lambda = |\mu |^2 \geq 0 .
\label{m5.11}
\end{equation}
This is clearly separable though not of product form. For the Mandel matrix
analysis, $| \mu \rangle \langle \mu |$ and $\hat{\rho}^{(ab)}_{D}$ are
equivalent. 

The matrices $C$, $C^{\dagger}$, $B$, $\Gamma$ involved in
$M^{(2)}(\hat{\rho}^{(ab)}_{D})$ are easy to calculate since
\begin{equation}
\langle \hat{a}^{\dagger} \hat{a} \rangle = \langle \hat{b}^{\dagger} \hat{b} \rangle
= \lambda, \,\,\langle \hat{a}^{\dagger 2} \hat{a}^2 \rangle =
\langle \hat{b}^{\dagger 2} \hat{b}^2 \rangle = {\lambda}^2.
\label{m5.12}
\end{equation}
We thus have\,:
\begin{eqnarray}
&& C = \left( \begin{array}{c}
\lambda \\ 0 \\ 0 \\ \lambda
\end{array} \right),\,\, \,\, 
B = \left( \begin{array}{cccc}
{\lambda}^{2} & 0 & 0 & {\lambda}^{2} + \lambda \\
0 & {\lambda}^{2} + \lambda & 0 & 0 \\
0 & 0 & {\lambda}^{2} + \lambda & 0 \\
{\lambda}^{2} + \lambda & 0 & 0 &{\lambda}^{2} 
\end{array} \right) ; \nonumber \\
&& \Gamma =
\left( \begin{array}{cccc}
0 & 0 & 0 & \lambda \\
0 & {\lambda}^{2} + \lambda & 0 & 0 \\
0 & 0 & {\lambda}^{2} + \lambda & 0 \\
\lambda & 0 & 0 & 0
\end{array} \right).
\label{m5.13}
\end{eqnarray}
The eigenvalues of $\Gamma$ being $\lambda(\lambda +1)$, $\lambda(\lambda +1)$,
$\lambda$, $-\lambda$, the state $\hat{\rho}^{(ab)}_{D}$ in (\ref{m5.11}) is
QO-noncl. To find its type we compute 
\begin{equation}
\psi_{0}(\alpha, \beta)^{\dagger} \Gamma \psi_{0}(\alpha, \beta) = 
2 |\alpha|^2 |\beta|^2\lambda(\lambda +2) \geq 0, 
\label{m5.14}
\end{equation}
so these states display hidden or Type II sub-PS.

In passing we note that the state $\hat{\rho}^{(a)}_{D}$ of mode $a$ obtained from Eq. (\ref{m5.11})
by tracing over $b$ alone is
\begin{eqnarray}
&&\hat{\rho}^{(a)}_{D} = {\rm Tr}_{b} \hat{\rho}^{(ab)}_{D} =
e^{- \lambda} \sum_{n=0}^{\infty} \frac{{\lambda}^n}{n!} |n \rangle_a {}_a \langle n |,
\label{m5.15}
\end{eqnarray}
for which the diagonal weight $P(I_a)$ is
\begin{eqnarray}
P(I_a) = \delta (I_a - \lambda).
\label{m5.16}
\end{eqnarray}
Partial trace over $a$ gives exactly similar results for mode $b$. Thus both
$\hat{\rho}^{(a)}_{D}$ and $\hat{\rho}^{(b)}_{D}$ are QO-cl, with their PND's
coinciding exactly with that of a coherent state.

Now we pass the two-mode state $\hat{\rho}^{(ab)}_{D}$ of Eq. (\ref{m5.11}) through
the BS $\hat{U}_0$ of Eq. (\ref{m5.1}); the resulting $\hat{\rho}'\,^{(ab)}_{D}$ is
\begin{eqnarray}
\hat{\rho}'\,^{(ab)}_{D} &=& \hat{U}_0 \hat{\rho}^{(ab)}_{D} \hat{U}_{0}^{-1} \nonumber \\
& =&e^{-\lambda} \sum_{n=0}^{\infty} \left(\frac{\lambda}{4}\right)^n \frac{1}{{n!}^3}
(\hat{a}^{\dagger 2} - \hat{b}^{\dagger 2} )^n | 0, 0 \rangle \langle 0, 0 |
(\hat{a}^2 -\hat{b}^2)^n .
\label{m5.17}
\end{eqnarray}
To apply the NPT entanglement test based on Eq. (\ref{m5.9})
it is convenient to examine $V_0 \Omega V_{0}^{T}$. Combining Eqs. (\ref{m5.4}, \ref{m5.13})
we find the matrices $\Gamma'$, $Y'$ associated with $\hat{\rho}'\,^{(ab)}_{D}$ to be
\begin{eqnarray}
\Gamma' =  V_0 \Gamma V_{0}^{T} &=&
\left( \begin{array}{cccc}
\frac{1}{2}{\lambda}^{2} + \lambda  & 0 & 0 & -\frac{1}{2}{\lambda}^{2} \\
0 & \frac{1}{2}{\lambda}^{2} & -\frac{1}{2}{\lambda}^{2} - \lambda & 0 \\
0 & -\frac{1}{2}{\lambda}^{2} - \lambda & \frac{1}{2}{\lambda}^{2} & 0 \\
-\frac{1}{2}{\lambda}^{2} & 0 & 0 & \frac{1}{2}{\lambda}^{2}+ \lambda
\end{array} \right), \nonumber \\
Y' & = & 
\left( \begin{array}{cccc}
\lambda & 0 & 0 & 0 \\
0 & 0 & 0 & 0 \\
0 & 0 & 2 \lambda +1 & 0 \\
0 & 0 & 0 & \lambda
\end{array} \right).
\label{m5.18}
\end{eqnarray}
Therefore according to Eq. (\ref{m5.9}) we have to test the positivity or otherwise of
\begin{eqnarray}
V_0 \Omega V_{0}^{T} &=& V_0 \Gamma V_{0}^{T} + Y' \nonumber \\
&=&\left( \begin{array}{cccc}
\frac{1}{2}{\lambda}^{2} + 2 \lambda  & 0 & 0 & -\frac{1}{2}{\lambda}^{2} \\
0 & \frac{1}{2}{\lambda}^{2} & -\frac{1}{2}{\lambda}^{2} - \lambda & 0 \\
0 & -\frac{1}{2}{\lambda}^{2} - \lambda & \frac{1}{2}{\lambda}^{2} + 2 \lambda +1 & 0 \\
-\frac{1}{2}{\lambda}^{2} & 0 & 0 & \frac{1}{2}{\lambda}^{2}+ 2 \lambda
\end{array} \right)
\label{m5.19}
\end{eqnarray}
The $(2,3)$ submatrix here is indefinite as it has determinant
$-\frac{1}{2}{\lambda}^{2}$. {\em This establishes that
$\hat{\rho}'\,^{(ab)}_{D}$ of Eq. (\ref{m5.17}) is NPT entangled}. Qualitatively
speaking, even though $Y'$ in Eq. (\ref{m5.18}) is nonnegative, the total matrix
$V_0 \Omega {V}_{0}^{T}$ in Eq. (\ref{m5.19}) is indefinite, with $\Gamma'$
dominating $Y'$. {\em The emphasis here has been to show that the NPT
entanglement produced by BS action can indeed be witnessed by the test based on
Eq. (\ref{m5.9}), which goes beyond the examination of the Mandel matrix in a
precise manner.} 

We can characterize the NPT entanglement we have proved in this
example further. From the expression in Eq. (\ref{m5.17}), the terms for $n=0$
and $n=1$ are respectively\,:
\begin{eqnarray}
&&e^{-\lambda} | 0, 0\rangle \langle 0, 0 |,\,\,\,\,\,
\frac{\lambda}{2} e^{-\lambda} (|2,0 \rangle - |0, 2 \rangle)(\langle 2,0| -
\langle 0, 2|),
\label{m5.20}
\end{eqnarray}
giving the matrix elements 
\begin{eqnarray}
&&\!\!\!\!\!\!\!\!\!\!\!\!\! (\hat{\rho}'\,^{(ab)}_{D})_{00,00} = e^{-\lambda} ;\nonumber \\
&&\!\!\!\!\!\!\!\!\!\!\!\!\!(\hat{\rho}'\,^{(ab)}_{D})_{20,20} = (\hat{\rho}'\,^{(ab)}_{D})_{02,02} = -
(\hat{\rho}'\,^{(ab)}_{D})_{20,02} = -(\hat{\rho}'\,^{(ab)}_{D})_{02,20} = \frac{\lambda}{2} 
e^{-\lambda}.
\label{m5.21}
\end{eqnarray}
One also obtains from the $n=2$ term in Eq. (\ref{m5.17}) the matrix element
\begin{equation}
(\hat{\rho}'\,^{(ab)}_{D})_{22,22} = \frac{\lambda^2}{8}e^{-\lambda}.
\label{m5.22}
\end{equation}

To demonstrate distillability we follow the recipe of Ref.\cite{horodecki97},
and project $\hat{\rho}'\,^{(ab)}_{D}$ into the $2\times 2$ bipartite subspace
spanned by $|0\rangle,\,|2\rangle$ of the $a$-mode  and  $|0\rangle,\,|2\rangle$
for the $b$-mode. The resulting $2\times 2$ state written in the basis
$|0,0\rangle,\,|0,2\rangle,\,|2,\,0\rangle,\,|2,2\rangle$ reads 
\begin{equation}
\hat{\rho}'\,^{(ab)}_{D} \rightarrow 
\left(\,\hat{\rho}'\,^{(ab)}_{D}\,\right)_{2\times 2} \sim  
\left( \begin{array}{cccc}
1 &0 & 0 & 0\\
0 & \frac{\lambda}{2}& -\frac{\lambda}{2} & 0\\ 
0 & -\frac{\lambda}{2}& \frac{\lambda}{2} & 0\\ 
0 & 0 & 0 & \frac{\lambda^2}{8}
\end{array} \right).
\label{m5.23}
\end{equation}
That $\left(\,\hat{\rho}'\,^{(ab)}_{D}\,\right)_{2\times 2}$  is inseparable is
manifest, for  its partial transpose has negative determinant.  Indeed, the
partial transpose  has an eigenvector of the form $\alpha |0,0 \rangle + \beta
|2,2 \rangle$ with a negative eigenvalue. {\em This demonstrates that the NPT
entanglement of $\hat{\rho}'\,^{(ab)}_{D}$ in Eq. (\ref{m5.17}) is  in fact
distillable}. 

\subsection{The two-mode squeezed vacuum}
This state defined in Eq. (\ref{m4.23}) is both pure and of product
form, i.e., of Schmidt rank one. It was studied in the previous
Section via its Mandel matrix, which showed it to be QO-noncl with
Type-II (i.e., hidden) sub-PS. In principle, after passing this state
through the BS $U_0$, we can apply the test based on Eq. (\ref{m5.9})
to see if the output state is NPT entangled. However, this involves some amount
of algebra. Fortunately,  the fact that the output state of the BS is entangled
can be seen  in this case  simply by inspection and without any calculations: 
\begin{eqnarray}
&&\!\!\!\!\!\!\!\!\!\!\!\!\!\!\!\!\!\!\!\!\!\!\!\!\hat{U}_0|\psi^{(ab)}
(\omega, \omega') \rangle  =  
\hat{U}_0 {\rm exp} \{{\frac{1}{4} (\xi \hat{a}^{\dagger 2} - \xi^{*}
\hat{a}^2)} +{\frac{1}{4} (\xi' \hat{b}^{\dagger 2} - \xi'^{*}
\hat{b}^2)} \} \hat{U}_{0}^{-1} | 0 , 0\rangle \nonumber \\
&& =  {\rm exp} 
\{ \frac{1}{8} (\xi (\hat{a}^{\dagger} - 
\hat{b}^{\dagger})^2 - \xi^{*} (\hat{a} - \hat{b})^2) +
\frac{1}{8} (\xi' (\hat{a}^{\dagger} + \hat{b}^{\dagger})^2 -
\xi'^{*} (\hat{a} + \hat{b})^2) \} | 0,0 \rangle\!.
\label{m5.24}
\end{eqnarray}
This is because the final unitary operator acting on $|0, 0 \rangle$ is clearly
not the tensor product of individual unitary operators acting separately on the
two vacua. It is of course important that  at least one of the factors $|
\psi^{(a)}(\omega)\rangle$, $|\psi^{(b)}(\omega')\rangle$ in the
initial product (\ref{m4.23}) {\bf be }  QO-noncl. A two-mode pure
product QO-cl state is necessarily a product of single-mode coherent
states, and  this  product structure is maintained by BS.

\section{Two-mode nonclassicality to three-mode entanglement}
In the preceding Section we studied the possibility of a $U(2)$ BS
converting a two-mode QO-noncl separable state into an entangled one
since both nonclassicality and entanglement are meaningful concepts
for such systems. The entire discussion was within the framework of
the space of states of a two-mode system.

Now we present a treatment of two-mode states analogous to that in Section II
for single-mode systems. That is, we couple a given two-mode state
$\hat{\rho}_{D}^{(ab)}$ to a third  ancilla mode in vacuum, pass such an input
state $\hat{\rho}^{(abc)}_{\rm in}$ through a `$U(3)$ beamsplitter'  (a
classicality preserving passive system), and obtain a three-mode output state
$\hat{\rho}_{\rm out}^{(abc)}$. We then test whether this shows NPT entanglement
as a consequence of (Mandel level) QO nonclassicality assumed to be present
initially in $\hat{\rho}_{D}^{(ab)}$; the PT operation is applied to the
$c$-mode. The motivation is to explore the algebraic expressions and form of the
test one is led to, apart from carrying the physical process described in
Section II to the next higher level. 

We begin with $\hat{\rho}^{(ab)}_{D}$ for
which $M^{(2)}(\hat{\rho}^{(ab)}_{D})$ shows QO nonclassicality. With  the
ancilla  $c$-mode in vacuum we have an input three-mode state 
\begin{eqnarray}
&& \hat{\rho}^{(abc)}_{\rm in} = \hat{\rho}^{(ab)}_{D} \otimes | 0 \rangle_c {}_c \langle 0 |,
\label{m6.1}
\end{eqnarray} 
strictly analogous to Eq. (\ref{m2.15}). To a general matrix $u\in U(3)$ we
associate a passive `beamsplitter' which unitarily mixes the annihilation
operators of the three modes in a manner analogous to Eq. (\ref{m2.13}), now
conserving $\hat{N}_a +\hat{N}_b +\hat{N}_c$. In the three-mode Hilbert space
this BS $u$ acts through a unitary operator $\hat{U}$, and we have
\cite{zeilinger94, dutta94} 
\begin{eqnarray}
 u &=& 
\left( \begin{array}{ccc}
u_{11} &u_{12} &u_{13} \\
u_{21}&u_{22}&u_{23} \\
u_{31}&u_{32}&u_{33} 
\end{array} \right) \in U(3) \,\, \rightarrow \,\,\hat{U}: \,\, \hat{U}^{\dagger}\hat{U}
 = \hat{U} \hat{U}^{\dagger} = 1,
\nonumber \\
\hat{U} \left( \begin{array}{c}
\hat{a} \\ \hat{b} \\ \hat{c}
\end{array} \right) \hat{U}^{-1} &=&
u^{\dagger} 
\left( \begin{array}{c}
\hat{a} \\ \hat{b} \\ \hat{c}
\end{array} \right) ,\,\,\,\,\,
\hat{U} \left( \begin{array}{c}
\hat{a}^{\dagger} \\ \hat{b}^{\dagger} \\ \hat{c}^{\dagger}
\end{array} \right) \hat{U}^{-1}=
u^T\left( \begin{array}{c}
\hat{a}^{\dagger} \\ \hat{b}^{\dagger} \\ \hat{c}^{\dagger}
\end{array} \right), \nonumber \\
\hat{U}^{-1} \left( \begin{array}{c}
\hat{a} \\ \hat{b} \\ \hat{c}
\end{array} \right) \hat{U}^{} &=&
u^{} 
\left( \begin{array}{c}
\hat{a} \\ \hat{b} \\ \hat{c}
\end{array} \right) ,\,\,\,\,\,
\hat{U}^{-1} \left( \begin{array}{c}
\hat{a}^{\dagger} \\ \hat{b}^{\dagger} \\ \hat{c}^{\dagger}
\end{array} \right) \hat{U}^{}=
u^* \left( \begin{array}{c}
\hat{a}^{\dagger} \\ \hat{b}^{\dagger} \\ \hat{c}^{\dagger}
\end{array} \right), \nonumber \\
\hat{U}(\hat{N}_a +\hat{N}_b +\hat{N}_c) &=& (\hat{N}_a +\hat{N}_b +\hat{N}_c)\hat{U}. 
\label{m6.2}
\end{eqnarray}
Therefore upon passage through this BS the state in Eq. (\ref{m6.1}) changes to
\begin{equation}
\hat{\rho}^{(abc)}_{\rm out} = \hat{U} \hat{\rho}^{(abc)}_{\rm in} \hat{U}^{-1} = 
\hat{U} \{ \hat{\rho}^{(ab)}_{D} \otimes | 0 \rangle_c {}_c \langle 0 |\} \hat{U}^{-1}.
\label{m6.3}
\end{equation}
To test this output state for NPT entanglement, we apply the PT operation to the
$c$-mode and then evaluate the `expectation value' of a suitably chosen
hermitian nonnegative operator\,:
\begin{eqnarray}
A\,\, &=& \alpha_0 + \alpha_1 \hat{a}\hat{c} +\alpha_2 \hat{b}\hat{c} +
\alpha_3 \hat{a}^{\dagger} \hat{c}^{\dagger} +
\alpha_4 \hat{b}^{\dagger} \hat{c}^{\dagger} : \nonumber \\
&&~{\rm Tr}(\hat{\rho}^{(abc)PT}_{\rm out} A^{\dagger} A) = \alpha^{\dagger} X 
\alpha,
\nonumber \\
\begin{array}{c}
X 
\\
\\
\\
\\
\\
\end{array}
&
\begin{array}{c}
=
\\
\\
\\
\\
\\
\end{array}
& 
\begin{array}{c}
{\rm Tr}(\hat{\rho}^{(abc)PT}_{\rm out}
\\
\\
\\
\\
\\
\end{array}
\left( \begin{array}{c}
1 \\ \hat{a}^{\dagger} \hat{c}^{\dagger} \\ \hat{b}^{\dagger} \hat{c}^{\dagger} \\
\hat{a}\hat{c} \\\hat{b}\hat{c}
\end{array} \right)
\begin{array}{c}
\left( \begin{array}{ccccc}
1 & \hat{a}\hat{c} & \hat{b}\hat{c} &\hat{a}^{\dagger} \hat{c}^{\dagger} &
 \hat{b}^{\dagger} \hat{c}^{\dagger}
\end{array}\right) );
\\
\\
\\
\\
\\
\end{array}
\,\alpha  =
\left( \begin{array}{c}
\alpha_0 \\ \alpha_1 \\\alpha_2 \\ \alpha_3 \\ \alpha_4
\end{array} \right).\,\,\,\,\,\,\,\,\,\,\,\,\,\,
\label{m6.4}
\end{eqnarray}
In spirit the approach here is similar to that used in the preceding Section
leading to Eq. (\ref{m5.7}), though here we are dealing with the extension from
two to three modes. The $5 \times 5$ hermitian matrix $X$, constructed by taking
entrywise expectation values as defined, is designed to be related to the input
Mandel matrix $M^{(2)}(\hat{\rho}^{(ab)}_{D})$ but going beyond it in a well
defined way. Developing it we find\,: 
\begin{eqnarray}
&&\left\{ \left( \begin{array}{c}
1 \\ \hat{a}^{\dagger} \hat{c}^{\dagger} \\ \hat{b}^{\dagger} \hat{c}^{\dagger} \\
\hat{a}\hat{c} \\\hat{b}\hat{c}
\end{array} \right)
\begin{array}{c}
\left( \begin{array}{ccccc}
1 & \hat{a}\hat{c} & \hat{b}\hat{c} &\hat{a}^{\dagger} \hat{c}^{\dagger} &
 \hat{b}^{\dagger} \hat{c}^{\dagger}
\end{array} \right)
\\
\\
\\
\\ 
\end{array}
\right\}^{PT} = \nonumber \\
&& :\left( \begin{array}{c}
1 \\ \hat{a}^{\dagger} \hat{c}^{} \\ \hat{b}^{\dagger} \hat{c}^{} \\
\hat{c}^{\dagger}\hat{a} \\\hat{c}^{\dagger}\hat{b}
\end{array} \right)
\begin{array}{c}
\left( \begin{array}{ccccc}
1 & \hat{c}^{\dagger}\hat{a} & \hat{c}^{\dagger}\hat{b} &\hat{a}^{\dagger} \hat{c}^{} &
 \hat{b}^{\dagger} \hat{c}^{}
\end{array} \right)
\\
\\
\\
\\
\end{array}: \,\, +
\left(\begin{array}{ccc}
\begin{array}{ccccc}
0&&0&&0 \\
0&&0&&0 \\
0&&0&&0
\end{array}
&
\begin{array}{c}
\vdots \\
\vdots \\
\vdots
\end{array}
& 
\begin{array}{ccc}
0&&0 \\
0&&0 \\
0&&0
\end{array} \\
\begin{array}{ccc}
\cdots&\cdots & \cdots
\end{array} &&
\begin{array}{cc}
\cdots & \cdots 
\end{array}
\\
\begin{array}{ccccc}
0&&0&&0 \\
0&&0&&0
\end{array}
&
\begin{array}{c}
\vdots \\
\vdots
\end{array}
& \hat{Z}
\end{array} \right),
\nonumber \\
&& \hat{Z} = \left( \begin{array}{cc}
\hat{a}^{\dagger}\hat{a}+ \hat{c}^{\dagger}\hat{c}+1 & \hat{b}^{\dagger}\hat{a} \\
\hat{a}^{\dagger}\hat{b} & \hat{b}^{\dagger}\hat{b}+\hat{c}^{\dagger}\hat{c}+1 
\end{array} \right) \nonumber \\
&& ~~~~~~=\; : \left( \begin{array}{c}
\hat{a} \\ \hat{b}
\end{array} \right)
\begin{array}{c}
\left( \begin{array}{cc}
\hat{a}^{\dagger} & \hat{b}^{\dagger}
\end{array} \right)
\\
\\
\end{array}:
 \,\, + \,\,(1+\hat{c}^{\dagger}\hat{c}) 
\left( \begin{array}{cc}
1 & 0 \\
0& 1 
\end{array} \right).
\label{m6.5}
\end{eqnarray}
Use this in Eq. (\ref{m6.4}), implement the conjugation $\hat{U}^{-1} 
(\cdots) \hat{U}$ and use the fact that the $c$-mode is initially in
vacuum to get\,:
\begin{eqnarray}
\begin{array}{c} 
X
\\ 
\\ 
\\ 
\\ 
\end{array} 
& 
\begin{array}{c}
 = 
\\ 
\\ 
\\ 
\\ 
\end{array} 
& 
\begin{array}{c}
 {\rm Tr}( \hat{\rho}^{(ab)}_{D}: 
\\ 
\\ 
\\ 
\\ 
\end{array} 
\left( \begin{array}{c} 1 \\ \hat{a}'^{\dagger} \hat{c}'^{} \\ \hat{b}'^{\dagger} \hat{c}'^{} \\ \hat{c}'^{\dagger}\hat{a}' \\\hat{c}'^{\dagger}\hat{b}' \end{array} \right) \begin{array}{c} \left( \begin{array}{ccccc} 1 & \hat{c}'^{\dagger}\hat{a}' & \hat{c}'^{\dagger}\hat{b}' &\hat{a}'^{\dagger} \hat{c}'^{} & \hat{b}'^{\dagger} \hat{c}'^{} \end{array}\right):
\\ 
\\ 
\\
 \\ 
\end{array} \nonumber \\ 
&&~~~~~+ {\rm Tr}(\hat{\rho}^{(ab)}_{D}
\left(\begin{array}{ccc} 
\begin{array}{ccccc} 0&&0&&0 \\ 0&&0&&0 \\ 0&&0&&0
\end{array} & \begin{array}{c} \vdots \\ \vdots \\ \vdots \end{array} &
\begin{array}{ccc} 0&&0 \\ 0&&0 \\ 0&&0 \end{array} \\ \begin{array}{ccc}
\cdots&\cdots & \cdots \end{array} && \begin{array}{cc} \cdots & \cdots
\end{array} \\ \begin{array}{ccccc} 0&&0&&0 \\ 0&&0&&0 \end{array} &
\begin{array}{c} \vdots \\ \vdots \end{array} & \hat{Z}' \end{array} \right)),
\nonumber \\ \hat{Z}' &=& \,\,: \left( \begin{array}{c} \hat{a}' \\ \hat{b}'
\end{array} \right) \begin{array}{c} \left( \begin{array}{cc} \hat{a}'^{\dagger}
& \hat{b}'^{\dagger} \end{array} \right) \\ \\ \end{array}: \,\, +
\,\,(1+\hat{c}'^{\dagger}\hat{c}') \left( \begin{array}{cc} 1 & 0 \\ 0& 1
\end{array} \right), \nonumber \\ \left( \begin{array}{c} \hat{a}' \\ \hat{b}'
\\ \hat{c}' \end{array} \right) &=& \left( \begin{array}{cc} u_{11} &u_{12} \\
u_{21} &u_{22} \\ u_{31} &u_{32} \end{array} \right) \left( \begin{array}{c}
\hat{a} \\ \hat{b} \end{array} \right). \label{m6.6} \end{eqnarray}
The appearance of the extra terms $\hat{Z}$, $\hat{Z}'$ is a result of normal ordering similar to the appearance of $\hat{Y}$ in Eq. (\ref{m5.6}). One can now disentangle the $u$- dependences and express the result in terms of $M^{(2)}(\hat{\rho}^{(ab)}_{D})$ and an additional piece involving $C= {\rm Tr}(\hat{\rho}_{D}^{(ab)} \hat{C})$\,:
\begin{eqnarray}
X &=& W(u) M^{(2)}(\hat{\rho}^{(ab)}_{D}) W(u)^{\dagger} \,\, + \,\,
\left(\begin{array}{ccc} \begin{array}{ccccc} 0&&0&&0 \\ 0&&0&&0 \\ 0&&0&&0
\end{array} & \begin{array}{c} \vdots \\ \vdots \\ \vdots \end{array} &
\begin{array}{ccc} 0&&0 \\ 0&&0 \\ 0&&0 \end{array} \\ \begin{array}{ccc}
\cdots&\cdots & \cdots \end{array} && \begin{array}{cc} \cdots & \cdots
\end{array} \\ \begin{array}{ccccc} 0&&0&&0 \\ 0&&0&&0 \end{array} &
\begin{array}{c} \vdots \\ \vdots \end{array} & {Z}' \end{array} \right),
\nonumber \\ W(u)& = & \left( \begin{array}{ccccc} 1 & \begin{array}{c} \vdots
\end{array} & \begin{array}{cc}
0\,\,\,\,\,\,\,\,\,\,\,&\,\,\,\,\,\,\,\,\,\,\,\,0 \end{array} &  \vdots &
\begin{array}{cc} 0\,\,\,\,\,\,\,\,\,\,\, & \,\,\,\,\,\,\,\,\,\,\,\,0
\end{array} \\ \cdots & & \cdots \,\,\, \cdots\,\,\, \cdots \,\,\, \cdots &&
\cdots \,\,\, \cdots\,\,\, \cdots \,\,\, \cdots \\ \begin{array}{c} 0\\0
\end{array} & \begin{array}{c}  \vdots \\ \vdots \end{array}& \left(
\begin{array}{c} u_{11}^{*} \\ u_{21}^{*} \end{array}  \right) \begin{array}{c}
\left( \begin{array}{cc} u_{31}^{} & u_{32}^{} \end{array}  \right) \\ \\
\end{array} & \begin{array}{c} \vdots \\ \vdots \end{array} & \left(
\begin{array}{c} u_{12}^{*} \\ u_{22}^{*} \end{array}  \right) \begin{array}{c} 
\left( \begin{array}{cc} u_{31}^{} & u_{32}^{} \end{array}  \right) \\ \\
\end{array} \\ \cdots & & \cdots \,\,\, \cdots\,\,\, \cdots \,\,\, \cdots &&
\cdots \,\,\, \cdots\,\,\, \cdots \,\,\, \cdots \\ \begin{array}{c} 0\\0
\end{array} & \begin{array}{c} \vdots \\ \vdots \end{array} & u_{31}^{*} \left(
\begin{array}{cc} u_{11} &u_{12} \\ u_{21} & u_{22} \end{array} \right) &
\begin{array}{c} \vdots \\ \vdots   \end{array} & u_{32}^{*} \left(
\begin{array}{cc} u_{11} & u_{12} \\ u_{21} & u_{22} \end{array} \right)
\end{array} \right), \nonumber \\ Z' &=& {\rm Tr}(\hat{\rho}^{(ab)}_{D}
\hat{Z}'), \nonumber \\ \hat{Z}' &=& \left( \begin{array}{cc} u_{11} &u_{12} \\
u_{21} & u_{22} \end{array} \right): \left( \begin{array}{c} \hat{a}\\\hat{b}
\end{array} \right) \begin{array}{c} \left( \begin{array}{cc} \hat{a}^{\dagger}
& \hat{b}^{\dagger} \end{array} \right) \\ \\ \end{array}: \left(
\begin{array}{cc} u_{11}^{*} &u_{21}^{*} \\ u_{12}^{*} & u_{22}^{*} \end{array}
\right) \nonumber \\ && + (1 + (u_{31}^{*} \hat{a}^{\dagger} +u_{32}^{*}
\hat{b}^{\dagger}) (u_{31}^{} \hat{a}^{} +u_{32}^{} \hat{b}^{})) \left(
\begin{array}{cc} 1& 0 \\ 0&1 \end{array} \right). \label{m6.7} \end{eqnarray}

If this matrix $X$, dependent on $\hat{\rho}^{(ab)}_{D}$ and $u\in U(3)$, is
indefinite, $\hat{\rho}^{(abc)}_{\rm out}$ of Eq. (\ref{m6.3}) is NPT entangled.
For this to happen, as we have assumed, $\hat{\rho}^{(ab)}_{D}$ must be
QO-noncl, since the BS $\hat{U}$ would map any QO-cl input into similar output, 
the ancilla being in a QO-classical state (vacuum).  

\subsection{Two illustrative examples}

The first is a two-mode state with only a finite number of photons, so that its 
QO nonclassicality is a foregone conclusion:
 \begin{eqnarray}
\hat{\rho}^{(ab)}_{D} &=& p| 2,0 \rangle \langle 2, 0|
 +q| 1,1 \rangle \langle 1, 1| +r| 0,2 \rangle \langle 0, 2|, \nonumber \\
&& p,q,r \geq 0 \,\,\,\,\,\,\,\,\,\,\, p+q+r =1.
\label{m6.8}
\end{eqnarray}
This is separable, though not a product state. The only non vanishing expectation 
values needed to construct the Mandel matrix are
\begin{eqnarray}
\langle \hat{a}^{\dagger} \hat{a} \rangle &=& 2p + q , \,\,
 \langle \hat{b}^{\dagger} \hat{b} \rangle = 2r + q, \,\,
 \langle \hat{a}^{\dagger 2} \hat{a}^2 \rangle = 2p, \, \nonumber \\
\langle \hat{a}^{\dagger} \hat{a}\hat{b}^{\dagger} \hat{b}\rangle &=& q, \,\,
\langle \hat{b}^{\dagger 2} \hat{b}^2 \rangle = 2r.
\label{m6.9}
\end{eqnarray}
Therefore the Mandel matrix is 
\begin{equation}
M^{(2)}(\hat{\rho}^{(ab)}_{D}) = 
\left( \begin{array}{ccccc}
1 & q+2p & 0 & 0 & q+2r \\
q+2p & 2p & 0 & 0 & q \\
0&0&q&0&0\\
0&0&0&q&0\\
q+2r & q & 0 & 0& 2r
\end{array} \right).
\label{m6.10}
\end{equation}
The determinants of various nontrivial $2\times2$ submatrices, the one
nontrivial $3\times 3$ submatrix, and finally of
$M^{(2)}(\hat{\rho}^{(ab)}_{D})$ itself, are (indicating the submatrices by the
relevant rows and columns)\,:
\begin{eqnarray}
&&(1,2):\,\,2p-(q+2p)^2;\,\,\,(1,5):\,\,2r-(q+2r)^2;\,\,\,(2,5): \,\,4pr-q^2; \nonumber \\
&&(1,2,5):\,\,q^2-4pr; \,\,\,{\rm det}M^{(2)}(\hat{\rho}^{(ab)}_{D})=\,\,q^2(q^2-4pr). 
\label{m6.11}
\end{eqnarray}
One can easily visualize situations for which the $(1,2)$ and $(1,5)$
submatrices become indefinite, for instance $q$ close to unity and $p,r$ close
to zero. In any case, since the $(2,5)$ subdeterminant is opposite to the
$(1,2,5)$ subdeterminant in sign and also to the full determinant, the state in
Eq. (\ref{m6.8}) is always QO-noncl at the Mandel matrix level.

The type of sub-PS can be easily determined. From Eq. (\ref{m6.10}) we find the 
$4 \times 4$ matrix $\Gamma$ to be\,: 
\begin{eqnarray}
\Gamma &=& 
\left( \begin{array}{cccc}
\delta_a & 0 & 0 & q -(q+2p)(q+2r) \\
0 & q & 0 & 0 \\
0 & 0 & q & 0 \\
q -(q+2p)(q+2r) & 0 & 0 & \delta_b
\end{array} \right), \nonumber \\
&& \nonumber \\
\delta_a & = & 2p - (q + 2p)^2 \, , \,\,\,\,\,\,\, \delta_b  = 2r - (q+2r)^2.
\label{m6.12}
\end{eqnarray}
Therefore also
\begin{eqnarray}
\psi_0 (\alpha, \beta)^{\dagger} \Gamma \psi_0 (\alpha, \beta) &=& 2p|\alpha|^4 +
4q |\alpha|^2 |\beta|^2 + 2r|\beta|^4 \nonumber \\
&&- ((q+2p) |\alpha|^2 + (q+2r) |\beta|^2)^2.
\label{m6.13}
\end{eqnarray}
For $\alpha =1$, $\beta=0$ this is $\delta_a$; for  $\alpha =0$, $\beta=1$
it is $\delta_b$. We now consider $p$ running over its range $[0,1]$ in successive
portions and draw corresponding conclusions\,:
\begin{eqnarray}
p = 0:&&\,\,\,\,q=0\,\,\, \Rightarrow \,\,\, \delta_b = -2; \,\,\,q > 0\,\,\,
\Rightarrow \,\,\, \delta_a < 0; \nonumber \\
0 < p < \frac{1}{2}: && \,\,\,\,\delta_a > 0 \,\,\, \Rightarrow \,\,\,
2p - (p-r+1)^2 > 0 \,\,\, \Rightarrow \,\,\, \nonumber \\
&& (p-r)^2 + 1 -2r < 0 \,\,\, \Rightarrow \,\,\, 2r > 1 \,\,\, \Rightarrow \,\,\, 
\delta_b < 0; \nonumber \\
&&\,\,\,\delta_a = 0 \,\,\, \Rightarrow \,\,\, (p-r)^2 + 1 -2r = 0\,\,\,
\Rightarrow \,\,\, p \neq r,\,\,\, \nonumber \\
&&2r > 1 \,\,\,\Rightarrow \,\,\, \delta_b < 0;
\nonumber \\
p = \frac{1}{2} :&& \,\,\,\, q=0\,\,\ \Rightarrow \,\,\, p=r = \frac{1}{2}, \,\,\,
\delta_a = \delta_b = 0; \nonumber \\
&&\,\,\,q > 0 \,\,\, \Rightarrow \,\,\, \delta_a < 0; \nonumber \\
\frac{1}{2} < p  \leq 1: &&\,\,\, 2p > 1 \,\,\, \Rightarrow \,\,\, \delta_a <0.
\label{m6.14}
\end{eqnarray}
Thus in every situation except $p=r= \frac{1}{2}$, $q=0$, either $\delta_a$
or $\delta_b$ is negative. In this one exceptional case we find from Eq. (\ref{m6.13})\,:
\begin{eqnarray}
&& p = r = \frac{1}{2}, \,\, q =0:\,\,\,\psi_0 (\alpha, \beta)^{\dagger}
\Gamma \psi_0 (\alpha, \beta) = -2 |\alpha|^2 |\beta|^2,
\label{m6.15}
\end{eqnarray}
which is negative for $\alpha, \beta \neq 0$. This establishes that the state
(\ref{m6.8}) is of Type I sub-PS.

Now we couple this state to the third $c$-mode in vacuum, and pass it through a particular
$U(3)$ BS, namely a $50:50$ BS acting on the $b$ and $c$ modes alone.
The output state is calculated using Eq. (\ref{m6.3}), and to test
whether it is NPT entangled we need to calculate the matrix $X$ 
of Eq. (\ref{m6.7}) involving the Mandel matrix
term and the added $Z'$ term. The choice of $u\in U(3)$, the resulting $W(u)$, and the
two parts of $X$ are as follows\,:
\begin{eqnarray}
u &=&
\left( \begin{array}{ccc}
1 &0 &0 \\
0 & 1/\sqrt{2} & 1/\sqrt{2} \\
0& -1/\sqrt{2}& 1/\sqrt{2}
\end{array} \right) \,\, \in U(3); \nonumber \\
W(u) &=& 
\left( \begin{array}{ccccc}
1&0&0&0&0 \\
0&0&-1/\sqrt{2}&0&0 \\
0&0&0&0&-1/2 \\
0&0&0&-1/\sqrt{2} &0 \\
0&0&0&0&-1/2
\end{array} \right); \nonumber \\
W(u)M^{(2)}(\hat{\rho}^{(ab)}_{D})W(u)^{\dagger} &=& 
\left( \begin{array}{cccccc}
1&0&-r-q/2&&0&-r-q/2 \\
0&q/2&0&&0&0 \\
-r-q/2&0&r/2&&0&r/2 \\
&&&\cdot&\cdots&\cdots \\
0&0&0&\vdots&q/2 &0 \\
-r-q/2&0&r/2&\vdots&0&r/2
\end{array} \right); \nonumber \\
Z' &=& \left( \begin{array}{cc}
2p+3q/2 +r+1 &0 \\
0& q+2r +1
\end{array} \right).
\label{m6.16} 
\end{eqnarray}
The dotted lines indicate where the $2 \times 2$ block $Z'$ has to be inserted.
Leaving out the trivial second and fourth rows and columns as they do not couple
to any others, the determinants of the various $2\times 2$ submatrices and the
$3\times3$ submatrix in $X$ are\,:
\begin{eqnarray}
&&\!\!\!\!\!\!\!\!\!\!\!\!\!\!\!\!(1,3):\,\,r/2-(q/2+r)^2;\,\,\,(1,5): \,\, 5r/2+ q+1- (q/2 + r)^2; \,\,\,\nonumber \\
&&\!\!\!\!\!\!\!\!\!\!\!\!\!\!\!\!(3,5):\,\,r(q+ 2r +1)/2; \,\,\,(1,3,5):\,\,(q+2r+1)(r/2-(q/2+r)^2).
\label{m6.17}
\end{eqnarray}
Comparing these with Eqs. (\ref{m6.11}) we see: whenever the QO nonclassicality of
$\hat{\rho}^{(ab)}_{D}$ manifests itself in the $(1,5)$ submatrix of 
$M^{(2)}(\hat{\rho}^{(ab)}_{D})$
being indefinite, simultaneously the 3-mode state $\hat{\rho}^{(abc)}_{\rm out}$
displays NPT entanglement. If on the other hand the $(1,2)$ submatrix
of $M^{(2)}(\hat{\rho}^{(ab)}_{D})$ were indefinite, then by suitably altering the $U(3)$ element
$u$ in Eq. (\ref{m6.16}) we can again achieve NPT entanglement of $\hat{\rho}^{(abc)}_{\rm out}$.
In either event, we see how a $U(3)$ BS can produce NPT entanglement
starting from a two-mode nonclassical state (\ref{m6.8}), and how the
signatures go beyond the indefiniteness of
$M^{(2)}(\hat{\rho}_{D}^{(ab)})$ in a precise manner.

The second example to illustrate the ideas of this Section is similar in
structure to the example (\ref{m5.11}) of the preceding Section, but differs in
certain details. For a real nonnegative parameter $\eta$ we define the separable
state
\begin{eqnarray}
&&\hat{\rho}^{(ab)}_{D} = \frac{1}{C} \sum_{n=0}^{\infty} \frac{{\eta}^{2n}}{(2n)!}
|n,n\rangle \langle n,n|, 
\label{m6.18}
\end{eqnarray}
where $C= {\rm Cosh}\, \eta$, and later $S = {\rm Sinh}\, \eta$ and $t = {\rm
tanh} \,\eta$. The case $\eta=0$ corresponds to the two-mode vacuum, and so we
take $0 < \eta < \infty$. Using the elementary sums 
\begin{eqnarray}
&&\sum_{n=0}^{\infty}( n\,\, {\rm or}\,\,n^2) \frac{{\eta}^{2n}}{(2n)!}=
\frac{\eta}{2} S \,\,{\rm or} \,\,\frac{\eta}{4}(S+ \eta C),
\label{m6.19}
\end{eqnarray}
the nonzero expectation values needed for the Mandel matrix are\,:
\begin{eqnarray}
&&\langle \hat{a}^{\dagger} \hat{a}\rangle = \langle \hat{b}^{\dagger} \hat{b}\rangle=
\frac{\eta}{2} t; \nonumber \\
&&\langle \hat{a}^{\dagger 2} \hat{a}^2\rangle = 
\langle \hat{b}^{\dagger 2} \hat{b}^2\rangle = \frac{\eta}{4} (\eta-t); \nonumber \\
&&\langle \hat{a}^{\dagger} \hat{b}^{\dagger} \hat{a} \hat{b}\rangle = 
\frac{\eta}{4} (\eta + t).
\label{m6.20}
\end{eqnarray}
Therefore we find\,:
\begin{eqnarray}
&& M^{2}(\hat{\rho}^{ab}_{D}) =
\left( \begin{array}{ccccc}
1 & \frac{\eta t}{2} & 0 & 0 & \frac{\eta t}{2} \\
\frac{\eta t}{2} & \frac{\eta}{4} (\eta-t) &0 & 0 &\frac{\eta}{4} (\eta+t)\\
0& 0&\frac{\eta}{4} (\eta+t) &0 &0 \\
0&0&0&\frac{\eta}{4} (\eta+t) & 0 \\
\frac{\eta t}{2} & \frac{\eta}{4} (\eta+t)& 0 & 0&\frac{\eta}{4} (\eta-t)
\end{array} \right).  
\label{m6.21}
\end{eqnarray}
Leaving out the third and fourth rows and columns, the remaining $2 \times 2$ 
subdeterminants are\,:
\begin{eqnarray}
&&(1,2)\,\,{\rm and}\,\,(1,5):\,\,\frac{\eta}{4} (\frac{\eta}{C^2} -t); \,\,\,(2,5):\,\,
-\frac{{\eta}^3 t}{4}.
\label{m6.22}
\end{eqnarray}
The function $\frac{\eta}{C^2} -t$ decreases monotonically from $0$ to $-1$
as $\eta$ runs from zero to infinity. We see that the state (\ref{m6.18}) is QO-noncl
for all $\eta > 0$.
To determine its Type we compute $\Gamma$ and its `expectation value' in 
$\psi_0(\alpha, \beta)$\,:
\begin{eqnarray}
\Gamma & = &
\frac{\eta}{4} \left( \begin{array}{cccc}
\frac{\eta}{C^2} -t & 0 & 0 & \frac{\eta}{C^2} +t \\
0 & \eta +t & 0 & 0 \\
0 & 0 & \eta +t & 0 \\
\frac{\eta}{C^2} +t & 0 & 0 &\frac{\eta}{C^2} -t,
\end{array} \right), \nonumber \\
\psi_{0}(\alpha, \beta)^{\dagger} \Gamma \psi_{0} (\alpha, \beta) &=& 
\frac{\eta}{4} \{\frac{\eta}{C^2} -t + 2|\alpha|^2|\beta|^2(\eta + 3t) \}.
\label{m6.23}
\end{eqnarray}
At both $\alpha=1$, $\beta=0$ and $\alpha=0$, $\beta=1$ the last expression is
negative, so the state (\ref{m6.18}) is QO-noncl Type I sub-PS. In this context
we note that the single-mode state $\hat{\rho}^{(a)}$ obtained from
(\ref{m6.18}) by tracing over $b$ is 
\begin{eqnarray}
\hat{\rho}^{(a)}_{D} & = & \frac{1}{C} \sum_{n=0}^{\infty} \frac{{\eta}^{2n}}{(2n)!}
|n \rangle_a {}_a \langle n |,
\label{m6.24}
\end{eqnarray}
and this has the Mandel matrix and determinant
\begin{eqnarray}
M^{(1)}(\hat{\rho}^{(a)}_{D}) & =& 
\left( \begin{array}{cc}
1 & \frac{\eta t}{2} \\
\frac{\eta t}{2} & \frac{\eta}{4} (\eta -t)
\end{array} \right), \nonumber \\
{\rm det}M^{(1)}(\hat{\rho}^{(a)}_{D}) &=& \frac{\eta}{4} (\frac{\eta}{C^2} -t) < 0.
\label{m6.25}
\end{eqnarray}
The properties of $\hat{\rho}^{(b)}_{D}$ are identical. So in contrast to the
state (\ref{m5.11}), now both $\hat{\rho}^{(a)}_{D}$ and $\hat{\rho}^{(b)}_{D}$
are QO-noncl, accompanying the Type I nature of $\hat{\rho}^{(ab)}_{D}$.

We now apply the NPT entanglement test outlined in Eqs. (\ref{m6.4}, \ref{m6.6},
\ref{m6.7}). The necessary expressions are\,:
\begin{eqnarray}
W(u) M^{(2)}(\hat{\rho}^{(ab)}) W(u)^{\dagger} &=&  
\left( \begin{array}{ccccc}
1 & 0 & -\frac{\eta t}{4} & 0 & -\frac{\eta t}{4} \\
0 & \frac{\eta}{8} ( \eta +t) & 0 & 0 &0 \\
-\frac{\eta t}{4} & 0 & \frac{\eta}{16} (\eta -t) & 0 &\frac{\eta}{16} (\eta -t) \\
0 & 0 & 0&\frac{\eta}{8}( \eta +t) & 0 \\
-\frac{\eta t}{4} & 0 &\frac{\eta}{16} (\eta -t) & 0 & \frac{\eta}{16} (\eta -t) 
\end{array} \right) , \nonumber \\
&&Z' =  
\left( \begin{array}{cc}
1 + \frac{3 \eta t}{4} & 0 \\
0 & 1 + \frac{\eta t}{2}
\end{array} \right).
\label{m6.26}
\end{eqnarray}
The $2 \times 2$ matrix $Z'$, which is positive definite, has to be `added' at
the lower right hand corner of the $5 \times 5$ matrix, leading to $X$ of Eq.
(\ref{m6.7}). Then the positivity or otherwise of $X$ has to be examined.
However, even without taking account of $Z'$, and unaffected by $Z'$, the
$(1,3)$ subdeterminant of $X$ is $\frac{\eta}{16}(\frac{\eta}{C^2} - t)$, which
is negative. This establishes the NPT entanglement of $\hat{\rho}^{(abc)}_{\rm
out}$ in this example.

The considerations of this Section show that the scheme described in Section II,
elevating single mode QO-noncl states to the two-mode level and then allowing BS
action to produce NPT entanglement, generalizes to the next higher level. The
method of Mandel matrices is a practical way to see these processes in action. 

\section{Genuine tripartite entanglement} Now that the main methods of our
approach-signatures of nonclassicality at the Mandel matrix level, their
nontrivial extensions to signatures of (NPT) entanglement created by BS action
-- have been applied to several examples, we go on to consider some more subtle
features of entanglement. We will show via an example that in the three-mode
case the BS action on an initial two-mode nonclassical state can lead to genuine
residual tripartite entanglement. This is in the sense of \cite{ckw00}, whereby
the output is a tripartite state similar to the GHZ state \cite{zeilinger99}\,:
there is no bipartite entanglement when any one of the three modes is traced
away. 

We consider the state (\ref{m5.11}) studied in Section V, and subject it
to the treatment of Section VI. As we have seen, this (separable) state shows
Type II sub-PS. With $\hat{\rho}^{(ab)}_{D}$ as in (\ref{m5.11}), we pass the
state (\ref{m6.1}),
\begin{eqnarray}
{\hat{\rho}}^{(ab)}_{D} \otimes | 0 \rangle_{c} {}_c \langle 0| = 
e^{-\lambda} \sum_{n=0}^{\infty} \frac{\lambda^n}{n!} |n, n \rangle_{ab}\,
{}_{ab}\langle n,n | \otimes | 0 \rangle_{c} {}_c \langle 0|,
\label{m7.1}
\end{eqnarray}
through a 50:50 $b-c$ BS, a special $U(3)$ element, whose action on the mode operators 
$\hat{b}$ and $\hat{c}$ is 
\begin{eqnarray}
\hat{U} \left(
\begin{array}{c}
\hat{b} \\
\hat{c}
\end{array}
\right) \hat{U}^{-1} =
\frac{1}{\sqrt{2}}
\left(
\begin{array}{cc}
1 &1 \\
-1 & 1
\end{array}
\right)
\left(
\begin{array}{c}
\hat{b} \\
\hat{c}
\end{array}
\right).
\label{m7.2}
\end{eqnarray}
The resulting state is
\begin{eqnarray}
{\hat{\rho}}^{(abc)}_{\rm out} &=&
\hat{U} ({\hat{\rho}}^{(ab)}_{D} \otimes | 0 \rangle_{c} {}_c \langle 0|) \hat{U}^{-1} 
\nonumber \\
&=& e^{-\lambda} \sum_{n=0}^{\infty}\frac{\lambda^n}{2^n n!} | n\rangle_{a} 
{}_{a} \langle n|
\otimes
{(\hat{b}^{\dagger} + \hat{c}^{\dagger})}^{n} | 0,0 \rangle_{bc}\, {}_{bc}
\langle 0, 0| {(\hat{b} + \hat{c})}^n \nonumber \\
&=&e^{-\lambda} \sum_{n=0}^{\infty}\frac{\lambda^n n!}{2^n} | n\rangle_{a} {}_{a} 
\langle n| \otimes
\sum_{r,s=0}^{n} \frac{|r, n-r \rangle_{bc}\, {}_{bc} \langle s, n-s |}
{\sqrt{r!(n-r)! s! (n-s)!}}.
\label{m7.3}
\end{eqnarray}
Clearly this is separable in the
$a/bc$ cut. However it is entangled in both the $c/ab$ and
$b/ac$ cuts as we show below.
As a test for NPT entanglement in the $c/ab$ cut, we evaluate the 
`expectation value' of a suitably chosen positive operator on the partially
transposed output ${\hat{\rho}}^{(abc)PT}_{\rm out}$, the partial transpose
being applied on the $c$ mode. For the choice of operator
$\hat{A}^{\dagger}\hat{A}$ where
\begin{eqnarray}
\hat{A}= {\alpha}_0 + {\alpha}_1 \hat{b} \hat{c} + {\alpha}_2 \hat{a}^{\dagger} \hat{a}, 
\label{m7.4}
\end{eqnarray} 
a test for NPT entanglement would be to check for violation of positivity of
\begin{eqnarray}
{\rm Tr}({\hat{\rho}}^{(abc)PT}_{\rm out} \hat{A}^{\dagger} \hat{A}) &=&
{\rm Tr}({\hat{\rho}}^{(abc)}_{\rm out}{(\hat{A}^{\dagger}
  \hat{A})}^{PT}) \nonumber \\
&=& {\begin{array}{c}( \begin{array}{ccc} \alpha_{0}^{*}& \alpha_{1}^{*} &
\alpha_{2}^{*} \end{array})\\ \\ \end{array}} {\begin{array}{c} X \\ \\
\end{array}} \left( \begin{array}{c} \alpha_0 \\ \alpha_1 \\ \alpha_2
\end{array}\right), \nonumber \\ X &=& {\rm Tr}({\hat{\rho}}^{(abc)}_{\rm out}
\left( \begin{array}{ccc} 1 & \hat{b} \hat{c}^{\dagger} & \hat{a}^{\dagger}
\hat{a} \\ \hat{b}^{\dagger} \hat{c} & \hat{b}^{\dagger}
\hat{b}\hat{c}^{\dagger} \hat{c} & \hat{b}^{\dagger} \hat{c} \hat{a}^{\dagger}
\hat{a} \\ \hat{a}^{\dagger} \hat{a} & \hat{a}^{\dagger} \hat{a} \hat{b}
\hat{c}^{\dagger} & \hat{a}^{\dagger} \hat{a}\hat{a}^{\dagger} \hat{a}
\end{array} \right) ). \label{m7.5} 
\end{eqnarray}
Using the fact that initially the $c$-mode is in the vacuum, we find\,:
\begin{eqnarray}
X= \left(
\begin{array}{ccc}
1 & {\lambda}/{2} & \lambda \\
{\lambda}/2 & {\lambda}^2/4 &\lambda(\lambda+1)/2\\
\lambda & \lambda(\lambda + 1)/2 & \lambda(\lambda + 1)
\end{array} \right).
\label{m7.6}
\end{eqnarray}
As the (2,3) submatrix of $X$ has negative determinant,
${\hat{\rho}}^{(abc)}_{\rm out}$ is NPT entangled across the $c/ab$ cut. It is
easy to see that a similar test with the same choice of $\hat{A}$, except that
now the PT operation is applied to the $b$ mode, yields the conclusion that
${\hat{\rho}}^{(abc)}_{\rm out}$ is NPT entangled across the $b/ac$ cut. So we
find in this example bipartite entanglement in a tripartite setup, as a result
of BS action.

Now to show that the entanglement is genuine tripartite, `residual' in the 
sense of \cite{ckw00}, we have the following\,:

\begin{eqnarray}
&&{\hat{\rho}}^{(ab)}_{\rm out} = {\rm Tr}_{c}({\hat{\rho}}^{(abc)}_{\rm out})=
e^{-\lambda}\sum_{n=0}^{\infty} \frac{\lambda^n n!}{2^n} | n \rangle_{a} \,{}_{a} 
\langle n|
\sum_{r=0}^{n} \frac{ |r \rangle_{b}\, {}_{b} \langle r|}{r!(n-r)!},\nonumber \\
&&{\hat{\rho}}^{(ac)}_{\rm out} = {\rm Tr}_{b}({\hat{\rho}}^{(abc)}_{\rm out})=
e^{-\lambda}\sum_{n=0}^{\infty} \frac{\lambda^n n!}{2^n} | n \rangle_{a} \,{}_{a} 
\langle n|
\sum_{r=0}^{n} \frac{ |r \rangle_{c}\, {}_{c} \langle r|}{r!(n-r)!},\nonumber \\
&&{\hat{\rho}}^{(bc)}_{\rm out} = {\rm Tr}_{a}({\hat{\rho}}^{(abc)}_{\rm out})=
 e^{-\lambda}\sum_{n=0}^{\infty} \frac{\lambda^n n!}{2^n}
 \sum_{r,s=0}^{n} \frac{|r, n-r \rangle_{bc}\, {}_{bc} \langle s, n-s |}
{\sqrt{r!(n-r)! s! (n-s)!}}.
\label{m7.7}
\end{eqnarray} 
The first two are manifestly separable. It may not be obvious at first glance
that the third is also separable but a closer look shows that it can be written
in the form
\begin{eqnarray}
{\hat{\rho}}^{(bc)}_{\rm out}= 
e^{-\lambda} \hat{U} ( \sum_{n=0}^{\infty} \frac{\lambda^n}{n!} |n \rangle_{b}\,
{}_{b}\langle n | \otimes | 0 \rangle_{c} {}_c \langle 0|)\hat{U}^{-1},
\label{m7.8}
\end{eqnarray}
where $U$ is the 50:50 $b-c$ BS (\ref{m7.2}).  Note that $\sum_{n=0}^{\infty}
\frac{\lambda^n}{n!} |n \rangle_{b}{}_{b}\langle n | $, the state of the $b$-
mode at the input, is simply the phase averaged (coarse grained) version of the
coherent state $|\sqrt{\lambda}\rangle$.  Thus ${\hat{\rho}}^{(bc)}_{\rm out}$
is the outcome of a  classical state passed through a BS, so it is classical and
hence separable.

It is interesting that the feature of genuine tripartite entanglement is
reminiscent of Type II sub-PS for a two-mode state at the Mandel level, where
the nonclassicality never shows up at any single mode level. An interesting
question in this context is the possibility of extension of monogamy relations
to  this non-Gaussian case \cite{osborne06, adesso06}.

\section{Further properties of Mandel parameters and beamsplitters}
For one and two mode field states, we have used the $2 \times 2$ and $5 \times
5$ Mandel matrices respectively to classify the states in a physically useful
manner. It is convenient to also have suitably normalized single parameter --
`scalar' -- measures of nonclassicality defined in terms of the Mandel matrices.
In the two-mode case a useful requirement would be invariance of such measures
under BS action. 

We begin with the single mode case. Here the Mandel $Q$
parameter was defined {\bf in \cite{mandel79}} as
\begin{eqnarray}
Q&=& \frac{(\Delta\hat{N}_a)^2 - \langle \hat{N}_a \rangle}{\langle
  \hat{N}_a \rangle} \nonumber \\
&=&\frac{\langle \hat{a}^{\dagger 2}\hat{a}^2 \rangle -
  \langle\hat{a}^{\dagger} \hat{a} \rangle^2}{\langle
  \hat{a}^{\dagger} \hat{a}\rangle} \nonumber \\
&=& \frac{{\rm det} M^{(1)}(\hat{\rho}^{(a)}_{D})}{\langle
  \hat{a}^{\dagger} \hat{a}\rangle}.
\label{m8.1}
\end{eqnarray}
Here all expectation values are in the state $\hat{\rho}^{(a)}_{D}$, and Eq.
(\ref{m4.2}) has been used. From the classification (\ref{m4.4}), $Q > 0$ and $Q
< 0$ correspond respectively to super-PS and nonclassical sub-PS cases. This
parameter is bounded below by $-1$, which is a convenient normalization. For $Q
> 0$ there is no upper bound.

In attempting to generalize to two modes, as a first step we show that the
separation of QO-noncl states into Types I and II is BS action invariant. For
any $u$ $\in$ $U(2)$, from Eq. (\ref{m2.13}) and the direct product structure of
$\hat{C}$ in Eq. (\ref{m4.8}) we easily obtain\,:
\begin{eqnarray}
\hat{U}^{-1} \hat{C} \hat{U}&=& V \hat{C},\,\,\,\,
\hat{U}_{}^{-1} \hat{C}^{\dagger} \hat{U} = 
\hat{C}^{\dagger} V_{}^{\dagger}, \nonumber \\
\hat{U}_{}^{-1} \hat{B} \hat{U}_{} &=& V \hat{B} V_{}^{\dagger}, \nonumber \\
V &=& u^* \otimes u.
\label{m8.2}
\end{eqnarray}
Therefore, with $C$ and $B$ defined as in Eq. (\ref{m4.15}), under
general BS action we have\,:
\begin{eqnarray}
\hat{\rho}^{'(ab)}_{D} &=& \hat{U} \hat{\rho}^{(ab)}_{D}
\hat{U}^{-1}\Rightarrow C'=VC, \,\,\,B'=V B V^{\dagger}, \nonumber \\
\Gamma'&=&V \Gamma V^{\dagger}.
\label{m8.3}
\end{eqnarray}
Now in the QO-noncl family of states, corresponding to
$M^{(2)}(\hat{\rho}_{D}^{(ab)})\not\geq 0$ or equivalently to $\Gamma
\not\geq 0$, the further separation into Types I and II is given in
Eq. (\ref{m4.18}). Here it is the `expectation values' of $\Gamma$ in four
component column vectors $\psi_0(\alpha, \beta)$ that are relevant. However
these vectors too have a direct product structure (\ref{m4.12}), so they are
mapped into similar vectors under the above changes\,:
\begin{eqnarray}
V^{\dagger} \psi_0 (\alpha, \beta)&=& 
(u^T \otimes u^{\dagger}) 
{\begin{array}{c} \\ 
\left(\begin{array}{c}
\alpha \\ \beta \end{array}\right) \otimes
\left(\begin{array}{c}
\alpha^* \\ \beta^* \end{array}\right)
\end{array}}
=\psi_0(\alpha', \beta'),
\nonumber \\ 
\left(\begin{array}{c}
\alpha' \\ \beta' \end{array}\right)&=&
u^T \left(\begin{array}{c}
\alpha \\ \beta \end{array}\right).
\label{m8.4}
\end{eqnarray}
This proves that the separation into Types I and II is preserved under
BS action. More specifically, if a BS converts a QO-noncl separable
state of a definite Type into an NPT entangled state, this change occurs
within the subfamily of that Type.

Now we generalize (\ref{m8.1}) to the two mode case. Keeping the requirement of
BS action invariance in mind, we define the two-mode Mandel parameter as
\begin{eqnarray}
Q'= \frac{{\rm Tr}(\Gamma) -||\Gamma||}{2(\langle\hat{a}^{\dagger}\hat{a}
\rangle + \langle \hat{b}^{\dagger}\hat{b} \rangle)},
\label{m8.5}
\end{eqnarray}
Here $||\Gamma||$ is the trace norm of $\Gamma$, which for hermitian $\Gamma$ is
the sum of the absolute values of its eigenvalues. Therefore $Q'$ is simply the
sum of the negative eigenvalues of $\Gamma$ divided by the expectation value of
the total number operator $\hat{N}_a + \hat{N}_b$. From Eq. (\ref{m8.3}), the
invariance of $Q'$ under BS action is obvious.

The parameter $Q'$ vanishes for QO-cl states (as defined via the Mandel matrix),
and is strictly negative for QO-noncl states. For two-mode product Fock states,
for instance, $Q'= -1$. To follow the distinction between the two Types, we
combine Eqs. (\ref{m4.12}, {\ref{m8.1}}) to define the variable single mode
Mandel parameter
\begin{eqnarray}
Q(\alpha, \beta)&=&\frac{\langle \hat{A}^{\dagger 2}\hat{A}^2 \rangle -
  \langle\hat{A}^{\dagger} \hat{A} \rangle^2}{\langle
  \hat{A}^{\dagger} \hat{A}\rangle}, \nonumber \\
\hat{A}&=& \alpha \hat{a} + \beta \hat{b}.
\label{m8.6}
\end{eqnarray}  
The minimum value of $Q(\alpha, \beta)$ as $\alpha$, $\beta$ vary is also
useful\,:
\begin{eqnarray}
Q^{\rm min}= \min_{|\alpha|^2 + | \beta|^2 =1} Q(\alpha, \beta).
\label{m8.7}
\end{eqnarray}
From all the previous discussions we draw up a table of results characterizing
various two-mode states (always at the Mandel level) in Table \ref{table}.
\begin{table}
\begin{tabular}{|c|c|c|}
\hline
{Category} & {Definition} & {Description}\\
\hline
(i) & $Q^{\rm min}\geq 0$, $Q' =0$ & QO-cl \\
(iia) & $Q^{\rm min} \leq Q' < 0$ & QO-noncl Type I \\
(iib) & $Q' < Q^{\rm min} < 0$ & QO-noncl Type I \\
(iii) & $Q^{\rm min}\geq 0$, $Q' < 0$ & QO-noncl Type II\\ 
\hline
\end{tabular}
\caption{\label{table}}
\end{table}
For some of these, we have examples from previous Sections. All QO-cl
states, including the states (\ref{m4.20}) for ${\rm det}
M^{(1)}(\hat{\rho}^{(a)}_{D}) \geq 0$, come under category (i). On the other
hand, the states (\ref{m4.20}) for ${\rm det} M^{(1)}(\hat{\rho}^{(a)}_{D}) < 0$
belong to category (ii). The subclassification into (iia) and (iib) is subtle,
but simple examples of each can be provided. For category (iia) we consider the
product of a Fock state at the $a$ mode and a coherent state at the $b$ mode\,:
\begin{eqnarray}
|\psi \rangle = |n \rangle_a \otimes |z \rangle_b.
\label{m8.8}
\end{eqnarray}   
This QO-noncl separable. The matrix $\Gamma$ is diagonal,
\begin{eqnarray}
\Gamma = {\rm diag}(-n, n|z|^2 , n|z|^2 , 0), 
\label{m8.9}
\end{eqnarray} 
leading to
\begin{eqnarray}
Q_2 = {-n}/(n+{|z|}^2) \geq -1. 
\label{m8.10}
\end{eqnarray}    
On the other hand, $Q(1,0)=-1$, so $Q^{\rm min} \leq Q' < 0$ which
falls under (iia). For category (iib) we can take the states
(\ref{m6.18}) which are QO-noncl and separable. Using
Eqs. (\ref{m6.20}, \ref{m6.23}) we find\,:
\begin{eqnarray}
Q'&=& -{1}/{2} \,: \nonumber \\
Q(\alpha, \beta) &=& \psi_0(\alpha, \beta)^{\dagger} \Gamma
\psi_0(\alpha, \beta)/\langle \hat{A}^{\dagger} \hat{A} \rangle
\nonumber \\
&=&\frac{1}{2t} \{\frac{\eta}{C^2} -t + 2|\alpha|^2|\beta|^2(\eta + 3t) \}. 
\label{m8.11}
\end{eqnarray} 
The minimum of $Q(\alpha, \beta)$ is reached when either $\alpha$ or
$\beta$ is zero\,:
\begin{eqnarray}
Q_{}^{\rm min} = \frac{1}{2t}\{\frac{\eta}{C^2} -t\} 
> - \frac{1}{2}.
\label{m8.12}
\end{eqnarray}
As $0 > Q^{\rm min} > Q'$, this falls under category (iib).

Another interesting example for (iib) is the
class of pure states obtained as an equal in-phase superposition
of product Fock states with given total occupation number $n$\,:
\begin{eqnarray}
|\psi_n \rangle = \frac{1}{\sqrt{n+1}} \sum_{r=0}^{n} |r,n-r \rangle.
\label{m8.13}
\end{eqnarray}
For the cases $n=1,2,3,4$ the numerically computed values of $Q'$ are $-1$,
$-1.085$, $-1.123$, $-1.143$ respectively. Since in any case $Q(\alpha, \beta)$
and $Q^{\rm min}$ are bounded below by $-1$, we have $Q' < Q^{\rm min}$. We may
also note that these entangled states cannot be produced by BS action on product
Fock states, except when $n=1$.

Turning finally to category (iii), we have examples from Sections 4 and 5. For
the two-mode squeezed vacuum state (\ref{m4.23}), we have using Eq.
(\ref{m4.26}) and the eigenvalue spectrum of $\Gamma$ stated after Eq.
(\ref{m4.25})\,: 
\begin{eqnarray}
Q' &=& SS'(SS'-CC')/(S^2 + S'^2) < 0, \nonumber \\
Q^{\rm min} &>& 0.
\label{m8.14}
\end{eqnarray}
For the family of states (\ref{m5.11}), from Eqs. (\ref{m5.12},
\ref{m5.13}) we find\,:
\begin{eqnarray}
Q'=- 1/2, \,\,\,\, Q^{\rm min} > 0.
\end{eqnarray}
So in both cases we have category (iii) states.

\section{conclusions} In this work we have investigated the relationships
between quantum optical nonclassicality of the phase insensitive type and
entanglement in multimode radiation fields In particular we have examined the
possibilities of converting nonclassicality in such fields into entanglement
through the use of  classicality preserving passive devices such as
beamsplitters. For the case of a single mode, after giving a complete
characterisation of the quantum optical nonclassicality at the level of phase
insensitive quantities, we have shown that such states through a beamsplitter
action with vacuum or more generally a coherent state at the other port, always
give rise to an NPT entangled state. For the case of two mode radiation fields
we have presented a test which  simultaneously witnesses both nonclassicality
and entanglement in such states and have also developed a scheme based on Mandel
matrices for characterising and classifying nonclassicality in one and two mode
states. In particular, it is shown that in the two mode case, the
characterisation at the level of Mandel matrices permits us to divide two mode
states into two categories--Type I where the nonclassicality  manifests itself
already at the single mode level and Type II where the nonclassicality is
intrinsically two mode in character with no Mandel type signatures of
nonclassicality at the single mode level. We have also examined in detail the
possibility of a $U(2)$ beamsplitter converting  Mandel level nonclassicality in
a two mode separable state into an NPT entangled state  and have given tests for
NPT entanglement in the state resulting from such an action. Distillability of
the state so produced is demonstrated in one case. Further, in a similar spirit
as for the case of a single mode, we have also analysed   the  action of a
$U(3)$ beamsplitter on a two mode Mandel level nonclassical state and have
derived conditions under which the two mode nonclassicality manifests itself in
NPT entanglement in the resulting three mode states. In this context, we have
also shown, via an example, how such an action can lead to a genuine tripartite
entangled state in the sense that there is no bipartite entanglement when any
one of the three modes is traced away. With a view to ease in categorisation of
nonclassicality in two mode states, by appealing to invariance under
beamsplitter action we have suggested analogues of the Mandel Q-parameter
originally introduced in the context of single mode radiation states.

\end{document}